\documentclass[USenglish]{article}	

\usepackage[utf8]{inputenc}				
\usepackage[big,online]{dgruyter}	
\usepackage{lmodern} 
\usepackage{microtype}
\usepackage[numbers,square,sort&compress]{natbib}
\usepackage{float}
\usepackage{anyfontsize}
\usepackage{subcaption}
\usepackage{soul}

\hypersetup{
    colorlinks=true,
    linkcolor=black,
    filecolor=black,
    citecolor = blue,
    urlcolor=cyan
}

\makeatletter
\newcommand{\owntag}[2][\relax]{
  \ifx#1\relax\relax\def\owntag@name{#2}\else\def\owntag@name{#1}\fi
  \refstepcounter{equation}\tag{Model \theequation, #2}%
  \expandafter\ltx@label\expandafter{eq:\owntag@name}%
  \edef\@currentlabel{\theequation, #2}\expandafter\ltx@label\expandafter{Eq:\owntag@name}%
  \def\@currentlabel{#2}\expandafter\ltx@label\expandafter{tag:\owntag@name}%
}
\makeatother

\theoremstyle{dgthm}

\theoremstyle{dgdef}

\begin{document}

	\articletype{Research Article}
	\received{Month	DD, YYYY}
	\revised{Month	DD, YYYY}
  \accepted{Month	DD, YYYY}
  \journalname{De~Gruyter~Journal}
  \journalyear{YYYY}
  \journalvolume{XX}
  \journalissue{X}
  \startpage{1}
  \aop
  \DOI{10.1515/sample-YYYY-XXXX}

\title{A comprehensive survey of the home advantage in American football}
\runningtitle{Home Advantage in American Football}


\author[1]{Luke Benz$^\dagger$}
\author[2]{Thompson Bliss$^\dagger$}
\author[3]{Michael Lopez} 
\runningauthor{Benz, Bliss, and Lopez}
\affil[1]{\protect\raggedright 
Department of Biostatistics, Harvard T.H. Chan School of Public Health, Boston MA  (lukebenz@g.harvard.edu)}
\affil[2]{\protect\raggedright 
National Football League, New York, NY (thompson.bliss@nfl.com)}
\affil[3]{\protect\raggedright 
National Football League, New York, NY (michael.lopez@nfl.com)}


\def\thefootnote{$^\dagger$}\footnotetext{These authors contributed equally to this work}\def\thefootnote{\arabic{footnote}}

	
\abstract{The existence and justification to the home advantage -- the benefit a sports team receives when playing at home -- has been studied across sport. The majority of research on this topic is limited to individual leagues in short time frames, which hinders extrapolation and a deeper understanding of possible causes. Using nearly two decades of data from the National Football League (NFL), the National Collegiate Athletic Association (NCAA), and high schools from across the United States, we provide a uniform approach to understanding the home advantage in American football. Our findings suggest home advantage is declining in the NFL and the highest levels of collegiate football, but not in amateur football. This increases the possibility that characteristics of the NCAA and NFL, such as travel improvements and instant replay, have helped level the playing field.}


\maketitle
\section{Introduction}\label{sec:intro}

Nearly every fan, player, and coach has tried to reconcile the impact of playing at home in football. In the National Football League (NFL) regular season, and in several high school playoff formats, teams fight tooth and nail for a home advantage (HA) in the postseason, in part because they've assumed it provides some form of a benefit over playing on the road. Likewise, home fans dress up and scream loudly when their team is on defense (and only when their team is on defense), in an effort to rankle the visitors. In the 2023 Wild Card round contest between Detroit and Los Angeles, for example, the noise made by the Detroit home crowd was estimated to reach 118 decibels, roughly the equivalent to an 737 airplane at takeoff \citep{kingFMIA}.  

The goal of our paper is to estimate the benefit to all that screaming. Currently, though arguably dwindling, the home advantage in football is considered to be on the margins of 2.5 points per year in the NFL and 3-5 points in National Collegiate Athletic Association (NCAA) play (see Section \ref{sec:lit_review}). However, most research into the home advantage is restricted to single leagues, small periods of time, or both single leagues over small periods of time. Additionally, several approaches assessing the home advantage fail to account for team strength, which can bias estimates, especially when better teams are more likely to play at home. Finally, the evolution of the game itself, including instant replay in college and in the NFL, and an increased emphasis on passing, has arguably changed the playing field for visitors. Home advantage has declined in other sports such as basketball (professional \cite{haberstroh2015, haberstroh2015b, kent2017} and collegiate \cite{davis2017, kenpom2017}), ice hockey \cite{jones2019}, and soccer \cite{lopesz2016, roeder2014}. As such, it is worth thoroughly assessing the likelihood of the home advantage declining in football, particularly in the larger context of these numerous factors that have changed in recent seasons, with the hope of better understanding possible drivers of home advantage.

Using three Bayesian models with differing assumptions, we (i) estimate the home advantage, (ii) identify if the home advantage has changed over time, and (iii) apply uniformly across all levels of football, including each of the 50 states, each collegiate level (FBS, FCS, Division II, and Division III), and the NFL. Models are fit in Stan \cite{standocs}, an open-source software for Bayesian inference with Markov-chain Monte Carlo (MCMC) sampling, and compared using leave-one-out cross-validation \cite{vehtari2017practical}. 

We find that the most extreme home advantages in 2023 tend to exist collegiately, with FCS leading the way at 2.49 points/game. Additionally, in FBS, FCS, and Division II, as well as the NFL, a significant decline in the home advantage is more probable than not. On the other hand, for all but a few states, high school home advantage since 2004 has been stagnant, or perhaps increasing. Our code and findings are all provided at \url{https://github.com/ThompsonJamesBliss/comprehensive_survey_american_football_home_adv}, for researchers in other sports or fields to build upon.

The paper is laid out as follows: Section \ref{sec:lit_review} reviews research into American football home advantage, Section \ref{sec:methods} details our Bayesian framework, Section \ref{sec:data} reviews our data, Section \ref{sec:results} presents results, and finally Section \ref{sec:discussion} concludes and comments on plausible explanations for our findings.

\section{Reviewing the Home Advantage in American Football}\label{sec:lit_review}

Nearly a half century of research has attempted to estimate home advantages in American football. This has been done using various models, scales, and time periods. Findings are universal -- that it's better to play at home -- but the size and magnitude of that benefit can vary slightly. Table \ref{tab:lit_review} provides an overview of roughly 20 papers across various levels of football, including the league, seasons, statistical model, scale (score differential or home team win percentage), and result.  

\begin{table}[htb]
\centering
\begin{tabular}{cccccc}
\hline\noalign{\smallskip}
\textbf{Paper} & \textbf{League} & \textbf{Seasons} & \textbf{Method} &  \textbf{Scale} & \textbf{Result}   \\
\noalign{\smallskip}\hline\noalign{\smallskip}
\citet{higgs2021} & NFL &  2016-2019 & Bayesian Neg. Binomial & Score Differential & 8\% \\
\citet{lopez2018} & NFL & 2006-2016  & Bayesian State-Space & Home Win \% & 8.9\% \\
\citet{glickman2017} & NFL & 2006-2014 & Bayesian State-Space & Score Differential & 2.4 Points \\
\citet{jones2016} & NFL & 1995-2014 & Empirical Home Win \% & Home Win \% & 3-12\% \\
\citet{david2011} & NFL  & 2008-2010 & Neural Networks & Score Differential & 3 Points \\ 
\citet{pollard2015} & NFL  & 2006-2009 & Empirical Home Win \% & Home Win \% & 6.2\% \\ 
\citet{baker2013} & NFL & 2001-2008 & Markov Process & Scoring Intensity & 0.1416$^{**}$ \\
\citet{glickman1998} & NFL & 1998-2003 & Bayesian State-Space & Score Differential & 3.2 Points \\
\citet{boulier2003} & NFL & 1994-2000 & Probit Regression & Home Win \% & 12.7\% \\
\citet{steffani1980} & NFL & 1970-1978 & Empirical Scores & Score Differential & 2 Points \\
\citet{harville1980} & NFL & 1970-1977 & Linear Model & Score Differential & 2-2.4 Points \\
\noalign{\smallskip}\hline\noalign{\smallskip}
\citet{fullagar2019} & NCAA & 2013, 2016 & Linear Mixed Model & Score Differential & 5 Points\\
\citet{wang2011} & NCAA & 2008-2009 & Multi-level Model & Score Differential & 5.9 Points$^\oplus$ \\
\citet{pollard2015} & NCAA  &  2006-2009 & Empirical Home Win \%  & Home Win \% &  12.8\% \\ 
\citet{caudill2007} & NCAA$^*$ & 1974-2005  & Linear Probability Model & Home Win \% & 11.4-22.5\% \\
\citet{gajewski2006} & NCAA$^\dagger$ & 1996-2004  & Bayesian Piecewise LM & Score Differential & 3-4 Points \\
\citet{massey1997} & NCAA & 1996 & Linear Model & Score Differential & 3.6 Points \\
\citet{steffani1980} & NCAA & 1970-1978 & Empirical Scores & Score Differential & 3 Points \\
\citet{harville1977} & NCAA & 1975 & Linear Model & Score Differential & 3.5 Points \\
\noalign{\smallskip}\hline\noalign{\smallskip}
\citet{mccutcheon2015} & High School$^\ddagger$ & 1982 & Empirical Scores & Score Differential & 1-4 Points \\
\citet{harville1977} & High School$^{\dagger\dagger}$ & 1975 & Linear Model & Score Differential & --- \\
\noalign{\smallskip}\hline\noalign{\smallskip}
\multicolumn{6}{l}{$^*$SEC; $^\dagger$Big 12, Big 10, Pac 10; $^\ddagger$Virginia; $^{\dagger\dagger}$Ohio, theoretical results only;$^{**}$Hazard ratio scale;} \\
\multicolumn{6}{l}{$^\oplus$ \citet{wang2011} report 5.9 points but include an intercept term in their model of of -2.8 points} \\
\multicolumn{6}{l}{indicating a team would beat an equal strength opponent by 3.1 points at home}\\
\noalign{\smallskip}\hline\noalign{\smallskip}
\end{tabular}
\caption{Summary of literature examining home advantage in American football across various levels. Papers frequently report the home advantage on two distinct scales, probability of beating an equal caliber opponent at home, or the expected score differential of playing an equal caliber opponent at home. For papers reporting home advantage on the win percentage scale, we report the advantage above 50\%. For example, a home advantage of 10\% would imply teams have a 50 + 10 = 60\% chance of beating an equal caliber opponent at home. Note that while, \citet{higgs2021} report results on the score differential scare, they report a home advantage of 8\%, which equates to an expected score differential that is 8\% of league average scoring if a team were to play an equal caliber opponent at home.}
\label{tab:lit_review}	
\end{table}

On a point scale, National Football League home advantage has varied from roughly 2 to 2.5 points \cite{steffani1980, harville1980} in the 1970's, to 2.5 to 3.5 points in more modern game play \cite{glickman2017, david2011}. This point difference equates to home teams winning at a rate somewhere between 56\% and 63\% of games against an equal caliber opponent at home \cite{lopez2018, pollard2015, higgs2021, boulier2003}. Though their primary focus was on the impact of Covid-19, \citet{higgs2021} concluded that the NFL's home advantage had been declining for multiple years prior to the pandemic. In NCAA, results are more extreme, with point differentials as high as five to six points in recent seasons, \cite{fullagar2019,  wang2011, gajewski2006}, which equates to home teams winning between 60\% and 65\% of the time \citep{pollard2015}. 

Typical statistical models for estimating home advantages include Bayesian frameworks (Negative Binomial and State Space paired comparison models), probit regression, hierarchical models, and neural networks. \citet{higgs2021} compared three versions of Bayesian models, including Poisson, Negative Binomial, and Normal, and found that Negative Binomial and Normal distributions performed best. 

The mechanisms behind the home advantage, including crowd impact and officiating tendencies, travel and rest discrepancies, fan interaction, visiting team playing style, and comfort level for the home team, have long been debated \citep{schwartz1977home}. Several NFL findings tie back to officiating and replay. In the book Scorecasting, \citet{moskowitz2011scorecasting} identified a connection between the advent of the NFL's instant replay process and the drop in fumble recovery rates for the home team. The implication is that, with additional technology, subjective fumble recovery decisions were eliminated. On the penalty side, \citet{snyder2015consistency} studied several common NFL infractions, finding that, for example, the odds of defensive pass interference fouls were 18\% higher when the home team was on offense, after accounting for score, time remaining, and team. However, it is unclear if this result was due to players fouling more often or varying tendencies of officials. \citet{vergin1999no} found a significantly higher home advantage for home teams on Monday night, which corresponds to both additional attention on the game and, potentially, more crowd noise.

Travel has also been linked to home advantage in football. In two seasons of NCAA data, \citet{fullagar2019} identified links between larger crowds and larger home advantage, as well as longer travel and worse performance for the visiting team. In the NFL, \citet{nichols2014impact} found that NFL home teams received a boost in performance when the visiting team has traveled further, although the effect was too small to turn a profit in betting markets.

\section{Methods}\label{sec:methods}

\subsection{Modeling Football Outcomes}\label{sec:models}
American football has a unique scoring system scoring where the most common scoring results contribute 3 (field goal), 7 (touchdown and extra point), 8 (touchdown and 2-point conversion),  6 (touchdown without any extra point or 2-point conversion), or 2 (safety) points towards a team's overall score. Despite the fact that a team's score is a sum of these discrete set of numbers, the difference between two teams' scores in any given game reasonably follows a Normal distribution \cite{glickman2017}.

Let $Y_{ijkt}$ be the score differential in a game between team $i$ and team $j$ in league $k$ during year $t$. We assume that 
$$
Y_{ijkt} \sim N(\mu_{ijkt}, \sigma^2_{k})
$$

\noindent We consider the following three models, which differ in how home advantage is modeled.

\begin{align}
    \mu_{ijkt} &= \theta_{ikt} - \theta_{jkt} + \alpha_k \owntag[model1]{Constant HA}\\
        \mu_{ijkt} &= \theta_{ikt} - \theta_{jkt} + \beta_{0k} + \beta_{1k}(t - t_0) \owntag[model2]{Linear HA}\\
        \mu_{ijkt} &= \theta_{ikt} - \theta_{jkt} + \gamma_{kt} \owntag[model3]{Time-Varying HA}
\end{align}

\noindent In the above three models, $\theta_{ikt}$ and $\theta_{jkt}$ represent team strength parameters for teams $i$ and $j$ in season $t$, respectively. 

In Model \ref{eq:model1}, $\alpha_k$ is a league specific home advantage parameter that is constant over time. In Model \ref{eq:model2}, home advantage is modeled as a linear trend over time where $\beta_{0k}$ denotes the home advantage in league $k$ during year $t_0$, the earliest year examined, and $\beta_{1k}$ denotes the year rate of change in home advantage in points/year. Finally, in Model \ref{eq:model3}, home advantage is denoted by a league-season specific term $\gamma_{kt}$. These three models, which we refer to as constant HA, linear HA, and time-varying HA, respectively, represent increases in model flexibility going from Model \ref{eq:model1} to Model \ref{eq:model3}. The choice of a linear home advantage trend in Model \ref{eq:model2} is designed to capture trends where there have been small changes in HA over time. 

We chose to assume that team strength parameters ($\theta_{ikt}$) are independent season to season rather than pursuing a dynamic state-space model, as has been used when analyzing the NFL in isolation \cite{lopez2018, glickman1998, glickman2017}. This choice was made in part due to large amounts of roster turnover at the high school and college levels between seasons, and the fact that some teams play very few games in a given season at the high school level. Additionally, some high school teams appear, disappear, and then appear again (see Section \ref{sec:data} for more details).  More importantly, we are not interested in conducting inference on team strength. Rather, we are only interested in team strength insofar as properly accounting for team strength is necessary for accurately estimating HA term(s) of interest and adequately characterizing trends in home advantage over time. 

\subsection{Model Fits in Stan}\label{sec:STAN}
We used Stan \cite{standocs}, an open-source statistical software designed for Bayesian inference with MCMC sampling, for each league $k$, and each of the three model options outlined in Section \ref{sec:models}. We chose to utilize a Bayesian approach when performing inference for several reasons. Of primary interest was obtaining posterior distributions of the change in home advantage \cite{benz2023estimating}. No paper referenced in Table \ref{tab:lit_review} has assessed HA change probabilistically. Additionally, the Bayesian framework allows for more flexibility when building models compared to standard methods like ordinary least squares (OLS) regression, particularly in how team strength are estimated. Finally, the decision to adhere to a Bayesian paradigm aligns our model building framework with those of \citet{glickman1998} and \citet{lopez2018}, two of the seminal works on home advantage in the NFL.

Models \ref{eq:model1}, \ref{eq:model2} and \ref{eq:model3} were fit using the following prior distributions. These prior distributions are weakly-informative and do not impose any outside knowledge on parameter estimation. 

$$
\begin{aligned}
  \theta_{ikt} &\sim N(0, \zeta^2_{k})~~~\text{(Team Strengths)} \\
  \zeta_k &\sim \text{HalfNormal}(0,5^2)~~~\text{(Team Strength Variance)}  \\
  \sigma_k  &\sim \text{HalfNormal}(0,5^2) ~~~\text{(Score Differential Variance)} \\ \\
  \alpha_k &\sim \text{Normal}(0, \eta^2_k) ~~~\text{(Model \ref{eq:model1} Home Advantage)} \\
  \eta_k &\sim \text{HalfNormal}(0,5^2) ~~~\text{(Model \ref{eq:model1}  Home Advantage Variance)} \\ \\
\end{aligned}
$$

$$
\begin{aligned}
  \beta_{0k} &\sim \text{Normal}(0, \lambda_{0k}^2) ~~~\text{(Model \ref{eq:model2} Home Advantage Intercept)} \\
  \beta_{1k} &\sim \text{Normal}(0, \lambda_{1k}^2) ~~~\text{(Model \ref{eq:model2} Home Advantage Trend)} \\
  \lambda_{0k} &\sim \text{HalfNormal}(0,5^2) ~~~\text{(Model \ref{eq:model2}  Home Advantage Intercept Variance)} \\
  \lambda_{1k} &\sim \text{HalfNormal}(0,5^2) ~~~\text{(Model \ref{eq:model2}  Home Advantage Trend Variance)}  \\ \\
 \gamma_{kt} &\sim \text{Normal}(0, \tau^2_{k}) ~~~\text{(Model \ref{eq:model3} Home Advantage)} \\
  \tau_{k} &\sim \text{HalfNormal}(0,5^2) ~~~\text{(Model \ref{eq:model3}  Home Advantage Variance)} 
  \nonumber
\end{aligned}
$$

Models were fit using 4 parallel chains, each made up of 2000 iterations, and a burn in of 500 draws. To check for model convergence, we examined the $\widehat R$ statistic \citep{gelman1992, brooks1998} for each parameter. If $\widehat R$ statistics are near 1, that indicates convergence \citep{bda3}, which was the case for all parameters in our model. To check for the informativeness of a parameter's posterior distribution, we also examined effective sample size (ESS, \cite{bda3}), which uses the relative independence of draws to equate the posterior distribution to the level of precision achieved in a simple random sample. Tables summarizing $\hat R$ statistics and effective sample sizes are available in the Supplementary Materials.

\subsection{Posterior Probability of Decline}\label{sec:p_decline}
We assessed the likelihood that home advantage has declined by computing the posterior probability $P(\beta_{1k} < 0)$, where $\beta_{1k}$  represents the average annual  change in HA in points/year. If this posterior probability is close to 1, there is strong evidence to suggest the HA in a given league has been declining significantly (at least in the statistical sense)  in a linear fashion over a long period of time. If this posterior probability is close to 0 (analogously, $P(\beta_{1k} > 0)$ is close to 1) there is strong evidence to suggest the HA in a given league has been increasing significantly over a long period of time. On the other hand, more moderate values of $(\beta_{1k} < 0)$ suggest there is less evidence to favor Model \ref{eq:model2} to Model \ref{eq:model1}.

\subsection{Model Comparison}\label{sec:model_comp}
For more formal model comparison, we computed expected log pointwise predictive density (ELPD) estimated via the leave-one-out cross-validation (LOO) approach of Vehtari et al. \cite{vehtari2017practical}. Operationalizing this approach entails computing the log-likelihood of each observation $Y_{ijkt}$ under each posterior sample, and supplying the resulting $n_k \times m$ matrix to \texttt{loo()} \cite{loopkg} in \texttt{R}, where $n_k$ denotes the total number of games analyzed in league $k$ and $m$ denotes the number of posterior samples. The primary motivation for model comparison via ELPD is that we were able to obtain associated standard error estimates, which enable comparisons between the difference in ELPD between models relative to the size of the associated standard error.

\section{Data}\label{sec:data}

\begin{figure}
    \centering
    \includegraphics[width=0.8\linewidth]{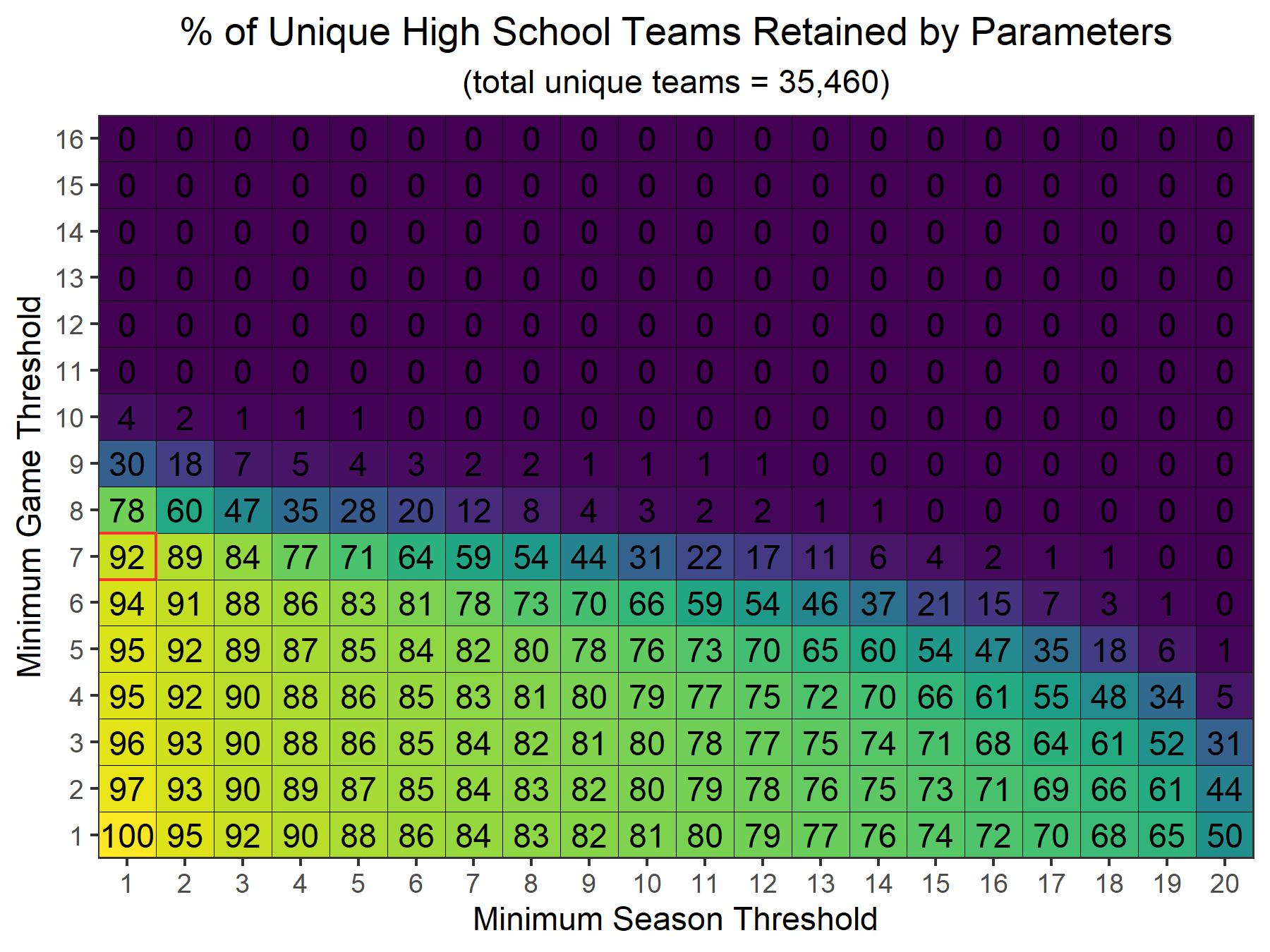}
    \caption{The percent of unique high school teams retained after an iterative filtering process for a given minimum games/seasons and minimum \# of seasons threshold. Teams that do not reach either the game or seasons threshold are removed from the data. Teams no longer reaching the threshold due to removal of opponents after the previous filtering are filtered out themselves. This process continues until no new teams are removed. The red outline indicates the resulting threshold choices for this work.}
    \label{fig:unique_teams}
\end{figure}

The data used for this analysis were comprised of games from the NFL, four divisions of NCAA, and all 50 US states. For ease, we refer to each of these entities as leagues, even though in practice there are several divisions within a single state for high school sports. Data collected include the season, the teams playing, the location (home or neutral site) and the scores of each team. NFL data were collected from an internal database (though such information is publicly available \cite{nflfastr}), while NCAA and high school data were scraped from MasseyRatings.com \cite{masseyratings} and Max Preps \cite{maxpreps} respectively. 

MasseyRatings.com is a website created by Kenneth Massey and has model-based team ratings for nearly every sport \cite{massey1997}. According to the website, data is collected electronically from a variety of publicly available domains with basic consistency checks run on multiple independent sources to verify the data's accuracy. Additionally, corrections and hard-to-find scores are entered manually. Max Preps is a news and data source for high school sports including information on players, teams and games. Game results are entered by a team's head coach and coaching staff.

We examined a nearly 20 year sample from 2004 - 2023, with the 2020 season excluded due to irregularities in play, travel, fan restrictions, and home advantage due to the COVID-19 pandemic \cite{higgs2021, benz2023estimating}. 2004 was chosen to be $t_0$ across all leagues as that was the earliest year Max Preps had public data available. Exceptions to $t_0 =$ 2004 were Alaska and North Dakota (2005), New Mexico (2006), and Wyoming (2007) where data that met the thresholds in Figure \ref{fig:unique_teams} weren't available until later years, as well as Oregon (2007) and Maryland (2008), where data quality was poor prior to the 2007 and 2008 season respectively. The high school data were filtered to include in-state games only, in order to more precisely measure the effect of home advantage in a particular state and remove travel effects that may arise from playing a a one-off game on the other side of the country.

One difficulty with the high school data was the presence of missing game results across seasons and teams. This is likely due inaccuracy from the crowd-sourced collection, and that some high school football programs have folded or combined with other school(s) during the 20 year period we chose to examine. Considering this missing data, it seemed ideal to remove teams with few observations, as a sufficient number of games for team $i$ in season $t$ is needed to properly estimate team strength parameter $\theta_{it}$. Of course, removing one team's game(s) from the dataset reduces the number of games for their opponents in the dataset as well. Thus, to reach our final high school sample, we applied the following iterative procedure.\\

\begin{enumerate}
    \item Pick thresholds for minimum \# of games per season and \# of seasons present in data.
    \item Remove team-seasons below these thresholds.
    \item Update \# of games/seasons a team appears in the data.
    \item Repeat steps 2 and 3 until no more teams are removed.
\end{enumerate}

\begin{table}[ht]
\centering
\begin{tabular}{ccc}
\hline
League & Number of Team/Seasons & Number of Games \\ 
\hline
NFL & 640 & 5,395 \\ 
  NCAA & 12,377 & 64,345 \\ 
  High School & 247,402 & 1,283,531 \\
   \hline
\end{tabular}
\caption{Counts of observations (games) and unique team/seasons by league.}
\label{tab:sample_size} 
\end{table}

Figure \ref{fig:unique_teams} displays the percentage of high school teams retained under an array of cutoffs for games/season and \# of seasons appearing in the data. As is apparent along the x-axis, it is quite uncommon for a team to have more than 2-3 observations across all 18 seasons of data. Additionally, it is quite uncommon to have greater than 7 observations in any given season. Ultimately, given the inconsistent coverage across seasons, we decided to treat each team-season independently without any prior data from previous iterations of the team informing the estimates, as in a dynamic state-space model \cite{glickman1998, glickman2017, lopez2018}. Given the extreme year-to-year roster turnover in high school relative to professional sports, and the fact that we are only interested in $\theta_{it}$ as a necessary adjustment towards estimating our suite of target home advantage parameters, this decision seems justifiable. 

After applying the iterative filtering algorithm, we restricted analysis to team-seasons with at least 7 games played, retaining 92\% of teams. Final sample sizes are available in Table \ref{tab:sample_size}.

\section{Results}\label{sec:results}

\subsection{Model Performance and Overall Home Advantage}

Results from Models \ref{eq:model1}, \ref{eq:model2}, and \ref{eq:model3} are shown in Table \ref{tab:elpd}. Of 55 leagues examined, the constant home advantage model (Model \ref{eq:model1}) had the best expected log predictive density in 29 leagues, the linear home advantage model (Model \ref{eq:model2}) had the best ELPD in 22 leagues, and the time-varying home advantage model (Model \ref{eq:model3}) was preferred in the remaining 6 leagues. Of note, Model \ref{eq:model2} was preferred for the NFL, and the top 3 NCAA (FBS, FCS, and Division II), but not in Division III. Out of 24 leagues where Model 1 did not have the best ELPD, none had an ELPD difference that was 4 standard errors worse than the preferred model, a common rule of thumb \citep{vehtari2020}, suggesting that differences from a constant HA model are small. This is not to say that home advantage is not changing, but rather the rate at which it is changing in small in comparison to the absolute magnitude of HA itself.

\begin{figure}[H]
    \centering
    \includegraphics[width = 0.8\textwidth]{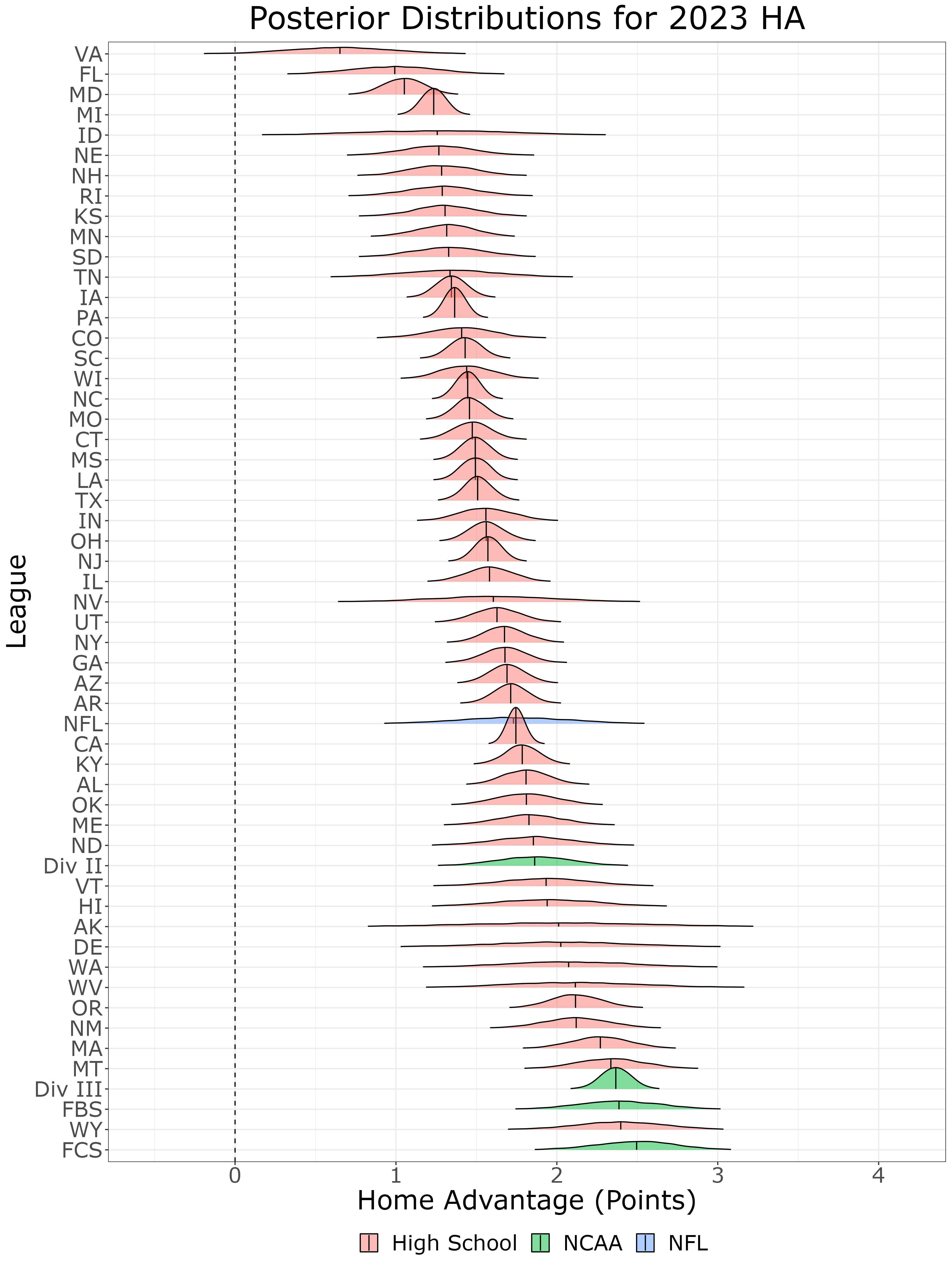}
    \caption{Posterior distributions of 2023 HA estimates from the best model for each league on the basis of ELPD. Posterior means and 95\% credible intervals are reported in Table \ref{tab:elpd}.}
    \label{fig:posteriors_2023}
\end{figure}

The estimated home advantage during the 2023 season from the best model in each respective state/league is shown in the last column in Table \ref{tab:elpd}. Posterior distributions of HA estimates for the 2023 season are displayed in Figure \ref{fig:posteriors_2023}. Across all 50 high school states, Wyoming had the greatest estimated home advantage in 2023, of 2.40 points (95\% credible interval 1.86, 2.93). Wyoming was one of only nine states where the home advantage was estimated to be 2 points or greater along with Alaska [2.01 (0.91, 3.12)], Delaware [2.01 (0.91, 3.12)], Massachusetts [2.27 (1.93, 2.61)], Montana [2.33 (1.93, 2.74)], New Mexico [2.12 (1.72, 2.52)], Pennsylvania [2.12 (1.82, 2.42)], Washington [2.08 (1.32, 2.84)], and West Virginia [2.12 (1.32, 3.00)]. At the other extreme, the lowest estimated home advantage in 2023 belonged to New Hampshire with an estimated HA of just 0.65 points (-0.01, 1.30), making it one of just two states with an estimated home advantage under 1 point, along with Florida [0.99 (0.47, 1.51)]. Other states with small home advantages included Maryland [1.05 (0.81, 1.29)], Michigan [1.23 (1.09, 1.37)], and Idaho [1.25 (0.30, 2.16)].

Relative to high school, estimated NCAA home advantages were higher. Greatest in 2023 was FCS [2.49 (1.99, 2.96)], followed by FBS [2.39 (1.89, 2.88)], Division III [2.37 (2.18, 2.55)], Division II [2.40 (2.20, 2.64)]. HA in all of the NCAA divisions were larger than our 2023 HA estimate for the NFL of 1.73 (1.07, 2.39).

\subsection{Trends in Home Advantage}

\begin{figure}[H]
    \centering
        \includegraphics[width = 0.8\textwidth]{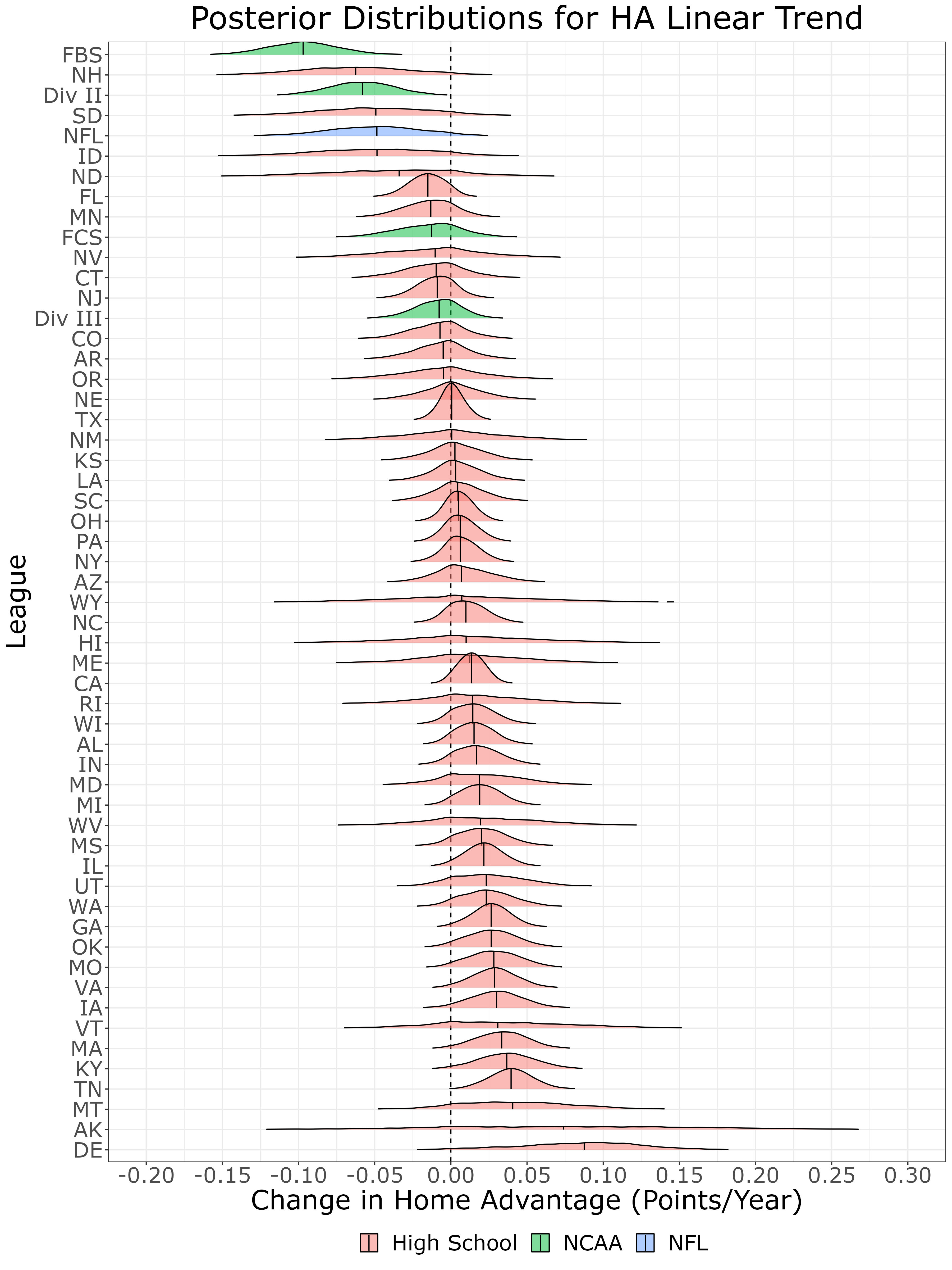}
    \caption{~\textbf{Left}: Posterior distributions for $\beta_{1k}$, the slope of the linear trend for HA in Model \ref{eq:model2}, which denotes the change in home advantage in points/year. Negative values of $\beta_{1k}$ denote a decline in HA while positive values of $\beta_{1k}$ denote an increase in HA.}
    \label{fig:posteriors}
\end{figure}

Figure \ref{fig:posteriors} displays posterior distribution for $\beta_{1k}$, the linear change in home advantage. Among the 55 leagues, FBS exhibited the largest estimated linear decline in HA, with a drop of roughly $\hat\beta_{1k} = -0.097$ points/year between 2004-2023 (roughly a point drop per decade). Related, the associated probability of decline was $P(\beta_{1k} < 0) = 1.000$. The other 3 NCAA divisions also had $\hat\beta_{1k} < 0$, though only Division II [$\hat\beta_{1k} = -0.058, P(\beta_{1k} < 0) = 0.999$] presented strong evidence of HA decline, while FCS [$\hat\beta_{1k} = -0.014, P(\beta_{1k} < 0) = 0.743$] and Division III [$\hat\beta_{1k} = -0.009, P(\beta_{1k} < 0) = 0.707$] exhibited less obvious linear HA changes over time.

Given both the preference of Model \ref{eq:model2} and $\hat\beta_{1k}$ which significantly differed from 0, there is evidence to suggest that while the two tiers of Division I football have similar 2023 estimated home advantages, the temporal trends in HA are distinct, with stronger evidence that home advantage in FBS has been on the decline for the past 2 decades than in FCS. Despite the fact that Model \ref{eq:model2} was preferred in the NFL, and the likelihood of a linear decline was deemed relatively high, with  $P(\beta_{1k} < 0) = 0.857$, such a trend is weaker than that observed in FBS, with $\hat\beta_{1k} = -0.032$ points/year (roughly 0.65 points in our twenty year sample). 

As observed in Figure \ref{fig:posteriors}, greater heterogeneity in home advantage trends was observed at the high school level with posterior means $\hat\beta_{1k}$ ranging from -0.063 points/year (New Hampshire) to 0.084 points/year (Delaware). While $\hat\beta_{1k} < 0$ in the NFL and all four NCAA divisions,  $\hat\beta_{1k} > 0$ in 38 of the 50 high school states and $< 0$ in the remaining 12. 

The probability of HA increase $P(\beta_{1k} > 0)$ exceeded 90\% in 14 states, albeit to varying degrees of practical significance. When fitting models on 50 states, however, we would expect a few significant states by chance. With this in mind, the most notable states with likely HA increase, on the basis of both statistical significance and practical significance were Delaware [$\hat\beta_{1k} = 0.084, P(\beta_{1k} > 0) = 0.968$], Tennessee [$\hat\beta_{1k} = 0.040, P(\beta_{1k} > 0) = 0.999$], Kentucky [$\hat\beta_{1k} = 0.038, P(\beta_{1k} > 0) = 0.977$] and Massachusetts [$\hat\beta_{1k} = 0.033, P(\beta_{1k} > 0) = 0.977$]. In the other direction,  New Hampshire [$\hat\beta_{1k} = -0.063, P(\beta_{1k} < 0) = 0.956$] and South Dakota $\hat\beta_{1k} = -0.049, P(\beta_{1k} < 0) = 0.908$] were noteworthy examples of states where HA has potentially been in decline.

\subsection{Visualizing Model Results}

\begin{figure}[H]
    \centering
    \includegraphics[width = 0.95\textwidth]{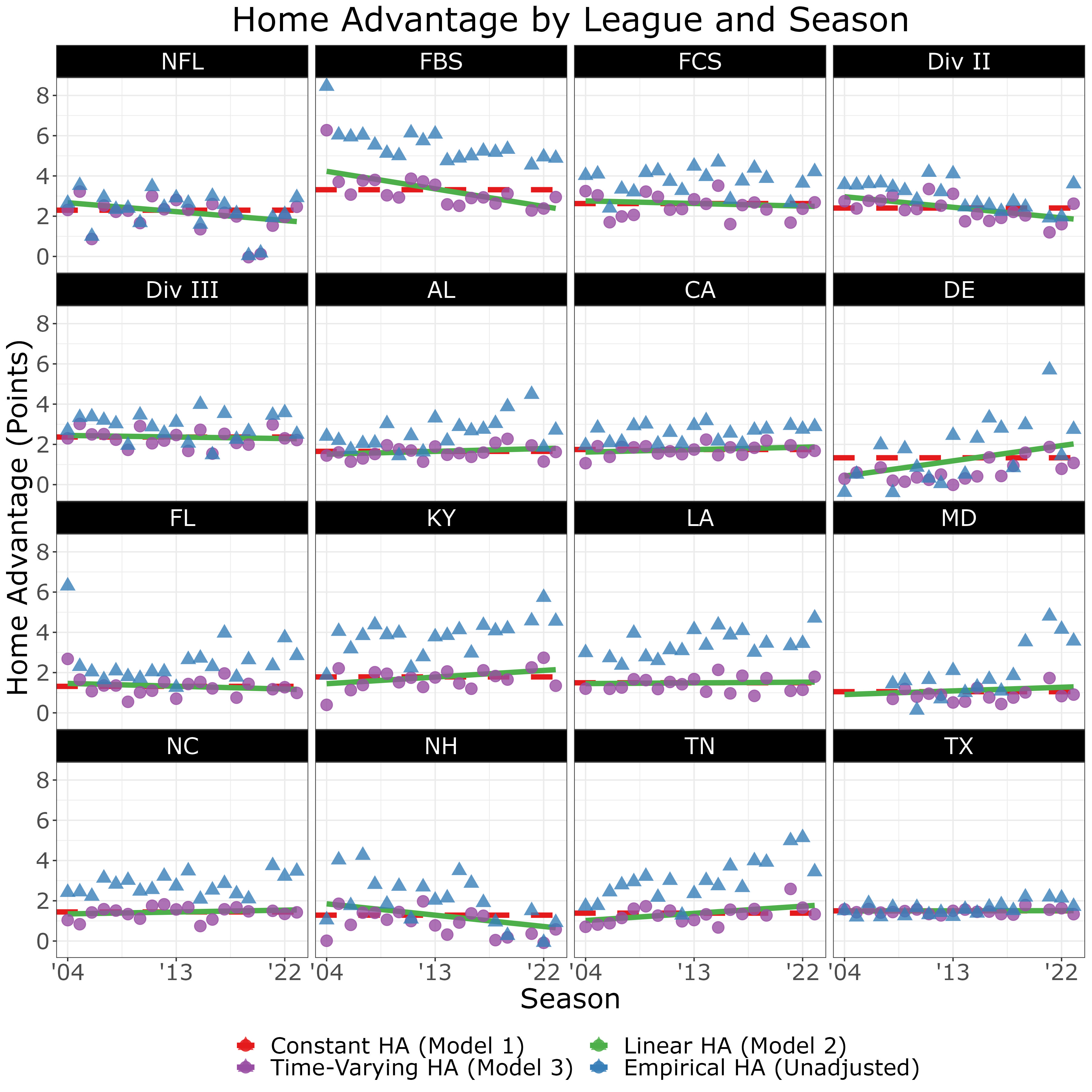}
    \caption{~Posterior mean home advantage over for select leagues. In addition to model-based estimates $\hat \alpha_{k}$ (Model \ref{eq:model1}), $\hat \beta_{0k} + \hat\beta_{1k}(t - t_0)$ (Model \ref{eq:model2}), and $\hat\gamma_{kt}$ (Model \ref{eq:model3}), we include $\hat \gamma^{\text{empirical}}_{kt}$, the mean home minus away score differential in season $t$ for league $k$, unadjusted for team strength. The leagues were selected to reflect heterogeneity in their HA trends over time. Of note is the big difference between $\hat \gamma^{\text{empirical}}_{kt}$ and model-based estimates in the majority of leagues, particularly in high school and the NCAA, where better teams are more likely to host home games.}
    \label{fig:ha_graphic}
\end{figure}

Figure \ref{fig:ha_graphic} displays model-based estimates of home advantage over time for the NFL, the four NCAA divisions, and select high school states. For comparison, we also present $\hat\gamma^{\text{empirical}}_{kt}$, the mean home minus away score-differential of all games in league $k$ during season $t$, unadjusted for team strength. This estimator is of interest as a point of comparison due to its use (or its analogue, empirical home win percentage) in previous works \cite{jones2016, pollard2015, steffani1980, mccutcheon2015}. 

One notable result is the difference between model-based home advantage estimates and unadjusted empirical estimates, as $\hat\gamma^{\text{empirical}}_{kt}$ nearly always exceeded the value of the analogous time-varying home advantage estimates, $\hat\gamma_{kt}$. This is likely because better teams more frequently play at home; in several high school playoff formats, better teams host playoff games, and in the NCAA, top programs pay worse opponents to visit and get blown out. 

In many cases differences between $\hat\gamma^{\text{empirical}}_{kt}$ and $\hat\gamma_{kt}$ exceeded 3 points, which itself was larger than the model-based home advantage estimates in every league. In some states, such as Louisiana, North Carolina, and Texas, $\hat\gamma^{\text{empirical}}_{kt}$ increased over time, while $\hat\gamma_{kt}$ remained fairly constant. This was further reflected by the fact that for each of these states $P(\beta_{1k} > 0)$ was not close to 1: Louisiana (0.593), North Carolina (0.794), Texas (0.535).

In Figure \ref{fig:ha_graphic}, Texas is an outlier amongst high school states in that there is very little difference between empirical and model-based home advantage estimates. The likely reason for this is that in Texas, playoff games after the first round are conducted at neutral sites equidistant from the two contending schools \cite{txrules}. Most other states, however, feature a playoff system where games are hosted by stronger teams. When this occurs, empirical estimates can be biased, by attributing part of a game's outcome to the home advantage, when it fact it's due to better teams playing at home.

\begin{table}[H]
\centering
\begin{tabular}{c|ccc|ccc|ccc|c}
  \hline
\multicolumn{1}{c|}{} & \multicolumn{3}{c|}{Model \ref{eq:model1} (Constant HA)} & \multicolumn{3}{c|}{Model \ref{eq:model2} (Linear HA)} & \multicolumn{3}{c|}{Model \ref{eq:model3} (Time-Varying HA)} & \multicolumn{1}{c}{}\\
League & $\Delta$ELPD & SE & \# SE &  $\Delta$ELPD &  SE & \# SE &  $\Delta$ELPD & SE & \# SE & 2023 HA  \\ 
  \hline
NFL & -1.31 & 1.75 & 0.75 & \textit{0.00} & \textit{0.00} & \textit{0.00} & -4.93 & 4.83 & 1.02 & 1.73 (1.07, 2.39) \\ 
\hline
  FBS & -11.03 & 4.75 & 2.32 & \textit{0.00} & \textit{0.00} & \textit{0.00} & -2.47 & 5.34 & 0.46 & 2.39 (1.89, 2.88) \\ 
  FCS & -0.30 & 1.63 & 0.19 & \textit{0.00} & \textit{0.00} & \textit{0.00} & -7.34 & 4.61 & 1.59 & 2.49 (1.99, 2.96) \\ 

  Div II & -3.90 & 3.56 & 1.09 & \textit{0.00} & \textit{0.00} & \textit{0.00} & -10.06 & 4.68 & 2.15 & 1.86 (1.41, 2.31) \\ 
  Div III & \textit{0.00} & \textit{0.00} & \textit{0.00} & -5.45 & 2.28 & 2.39 & -13.11 & 5.52 & 2.38 & 2.37 (2.18, 2.55) \\ 
\hline
  AK & \textit{0.00} & \textit{0.00} & \textit{0.00} & -0.04 & 0.99 & 0.04 & -1.50 & 3.08 & 0.49 & 2.01 (0.91, 3.12) \\ 
  AL & -0.68 & 3.55 & 0.19 & \textit{0.00} & \textit{0.00} & \textit{0.00} & -14.66 & 6.76 & 2.17 & 1.81 (1.54, 2.08) \\ 
  AR & \textit{0.00} & \textit{0.00} & \textit{0.00} & -6.15 & 2.05 & 3.00 & -13.25 & 4.61 & 2.87 & 1.71 (1.51, 1.92) \\ 
  AZ & \textit{0.00} & \textit{0.00} & \textit{0.00} & -2.55 & 2.21 & 1.15 & -12.94 & 5.37 & 2.41 & 1.69 (1.48, 1.90) \\ 
  CA & \textit{0.00} & \textit{0.00} & \textit{0.00} & -4.91 & 4.98 & 0.99 & -9.45 & 7.87 & 1.20 & 1.75 (1.65, 1.84) \\ 
  CO & -1.09 & 2.24 & 0.49 & \textit{0.00} & \textit{0.00} & \textit{0.00} & -16.07 & 4.92 & 3.27 & 1.40 (0.99, 1.80) \\ 
  CT & \textit{0.00} & \textit{0.00} & \textit{0.00} & -0.14 & 1.91 & 0.07 & -5.88 & 4.54 & 1.29 & 1.47 (1.25, 1.70) \\ 
  DE & -0.10 & 2.35 & 0.04 & \textit{0.00} & \textit{0.00} & \textit{0.00} & -7.98 & 3.24 & 2.46 & 2.02 (1.14, 2.88) \\ 
  FL & -6.67 & 8.38 & 0.80 & -1.66 & 8.10 & 0.21 & \textit{0.00} & \textit{0.00} & \textit{0.00} & 0.99 (0.47, 1.51) \\ 
  GA & -0.72 & 3.86 & 0.19 & \textit{0.00} & \textit{0.00} & \textit{0.00} & -5.77 & 5.79 & 1.00 & 1.68 (1.41, 1.93) \\ 
  HI & \textit{0.00} & \textit{0.00} & \textit{0.00} & -0.72 & 1.14 & 0.63 & -4.62 & 3.73 & 1.24 & 1.94 (1.34, 2.55) \\ 
  IA & \textit{0.00} & \textit{0.00} & \textit{0.00} & -2.57 & 3.16 & 0.81 & -16.58 & 5.03 & 3.29 & 1.34 (1.16, 1.52) \\ 
  ID & -0.23 & 1.68 & 0.14 & \textit{0.00} & \textit{0.00} & \textit{0.00} & -4.72 & 3.78 & 1.25 & 1.25 (0.30, 2.16) \\ 
  IL & -3.10 & 4.13 & 0.75 & \textit{0.00} & \textit{0.00} & \textit{0.00} & -11.32 & 5.58 & 2.03 & 1.58 (1.31, 1.85) \\ 
  IN & -4.66 & 3.00 & 1.56 & \textit{0.00} & \textit{0.00} & \textit{0.00} & -17.79 & 4.28 & 4.16 & 1.56 (1.26, 1.89) \\ 
  KS & -4.06 & 2.22 & 1.83 & \textit{0.00} & \textit{0.00} & \textit{0.00} & -6.12 & 4.92 & 1.24 & 1.31 (0.91, 1.70) \\ 
  KY & \textit{0.00} & \textit{0.00} & \textit{0.00} & -2.54 & 3.07 & 0.83 & -1.59 & 5.74 & 0.28 & 1.79 (1.58, 1.99) \\ 
  LA & \textit{0.00} & \textit{0.00} & \textit{0.00} & -2.32 & 2.33 & 1.00 & -7.96 & 5.12 & 1.56 & 1.49 (1.33, 1.66) \\ 
  MA & -2.12 & 3.14 & 0.68 & \textit{0.00} & \textit{0.00} & \textit{0.00} & -6.84 & 4.91 & 1.39 & 2.27 (1.93, 2.61) \\ 
  MD & \textit{0.00} & \textit{0.00} & \textit{0.00} & -2.80 & 2.31 & 1.21 & -2.50 & 4.01 & 0.62 & 1.05 (0.81, 1.29) \\ 
  ME & \textit{0.00} & \textit{0.00} & \textit{0.00} & -4.31 & 1.18 & 3.65 & -9.27 & 4.40 & 2.11 & 1.83 (1.43, 2.23) \\ 
  MI & \textit{0.00} & \textit{0.00} & \textit{0.00} & -3.33 & 3.75 & 0.89 & -2.96 & 6.02 & 0.49 & 1.23 (1.09, 1.37) \\ 
  MN & -2.39 & 2.60 & 0.92 & \textit{0.00} & \textit{0.00} & \textit{0.00} & -6.42 & 5.49 & 1.17 & 1.31 (0.96, 1.64) \\ 
  MO & \textit{0.00} & \textit{0.00} & \textit{0.00} & -3.45 & 3.06 & 1.13 & -16.10 & 4.84 & 3.32 & 1.46 (1.28, 1.64) \\ 
  MS & \textit{0.00} & \textit{0.00} & \textit{0.00} & -0.22 & 2.59 & 0.09 & -13.46 & 4.33 & 3.11 & 1.49 (1.33, 1.67) \\ 
  MT & \textit{0.00} & \textit{0.00} & \textit{0.00} & -0.27 & 1.71 & 0.16 & -6.46 & 4.79 & 1.35 & 2.33 (1.93, 2.74) \\ 
  NC & \textit{0.00} & \textit{0.00} & \textit{0.00} & -1.47 & 2.78 & 0.53 & -4.50 & 5.39 & 0.83 & 1.45 (1.31, 1.58) \\ 
  ND & \textit{0.00} & \textit{0.00} & \textit{0.00} & -0.45 & 1.40 & 0.32 & -9.02 & 3.76 & 2.40 & 1.85 (1.35, 2.34) \\ 
  NE & -1.40 & 1.93 & 0.72 & \textit{0.00} & \textit{0.00} & \textit{0.00} & -10.04 & 4.03 & 2.49 & 1.27 (0.83, 1.72) \\ 
  NH & \textit{0.00} & \textit{0.00} & \textit{0.00} & -0.50 & 2.24 & 0.22 & -8.70 & 4.00 & 2.18 & 1.29 (0.89, 1.67) \\ 
  NJ & \textit{0.00} & \textit{0.00} & \textit{0.00} & -0.97 & 2.80 & 0.34 & -10.76 & 5.29 & 2.03 & 1.57 (1.42, 1.72) \\ 
  NM & \textit{0.00} & \textit{0.00} & \textit{0.00} & -4.85 & 1.33 & 3.64 & -11.48 & 3.09 & 3.72 & 2.12 (1.72, 2.52) \\ 
  NV & -0.80 & 1.19 & 0.67 & \textit{0.00} & \textit{0.00} & \textit{0.00} & -15.01 & 3.72 & 4.04 & 1.61 (0.81, 2.37) \\ 
  NY & -3.95 & 3.67 & 1.08 & \textit{0.00} & \textit{0.00} & \textit{0.00} & -5.17 & 6.80 & 0.76 & 1.68 (1.43, 1.93) \\ 
  OH & -1.42 & 3.98 & 0.36 & \textit{0.00} & \textit{0.00} & \textit{0.00} & -10.35 & 5.69 & 1.82 & 1.56 (1.36, 1.77) \\ 
  OK & -1.73 & 2.94 & 0.59 & \textit{0.00} & \textit{0.00} & \textit{0.00} & -9.53 & 5.44 & 1.75 & 1.81 (1.46, 2.17) \\ 
  OR & \textit{0.00} & \textit{0.00} & \textit{0.00} & -0.76 & 1.55 & 0.49 & -10.19 & 3.60 & 2.83 & 2.12 (1.82, 2.42) \\ 
  PA & \textit{0.00} & \textit{0.00} & \textit{0.00} & -7.18 & 3.44 & 2.08 & -21.42 & 5.14 & 4.17 & 1.37 (1.25, 1.48) \\ 
  RI & \textit{0.00} & \textit{0.00} & \textit{0.00} & -1.58 & 0.97 & 1.62 & -11.64 & 3.48 & 3.35 & 1.28 (0.85, 1.71) \\ 
  SC & \textit{0.00} & \textit{0.00} & \textit{0.00} & -3.11 & 2.43 & 1.28 & -16.93 & 4.81 & 3.52 & 1.43 (1.25, 1.62) \\ 
  SD & \textit{0.00} & \textit{0.00} & \textit{0.00} & -1.30 & 2.84 & 0.46 & -6.93 & 4.59 & 1.51 & 1.33 (0.90, 1.74) \\ 
  TN & -3.45 & 6.65 & 0.52 & -1.79 & 5.89 & 0.30 & \textit{0.00} & \textit{0.00} & \textit{0.00} & 1.34 (0.74, 1.94) \\ 
  TX & -5.92 & 5.04 & 1.18 & \textit{0.00} & \textit{0.00} & \textit{0.00} & -29.86 & 6.12 & 4.88 & 1.51 (1.35, 1.68) \\ 
  UT & \textit{0.00} & \textit{0.00} & \textit{0.00} & -1.04 & 1.95 & 0.54 & -14.24 & 4.09 & 3.48 & 1.63 (1.35, 1.91) \\ 
  VA & -0.48 & 7.33 & 0.07 & -0.84 & 7.01 & 0.12 & \textit{0.00} & \textit{0.00} & \textit{0.00} & 0.65 (-0.01, 1.30) \\ 
  VT & \textit{0.00} & \textit{0.00} & \textit{0.00} & -0.03 & 1.13 & 0.02 & -9.74 & 3.84 & 2.54 & 1.93 (1.38, 2.47) \\ 
  WA & -0.66 & 6.54 & 0.10 & -3.53 & 6.35 & 0.56 & \textit{0.00} & \textit{0.00} & \textit{0.00} & 2.08 (1.32, 2.84) \\ 
  WI & -2.58 & 3.45 & 0.75 & \textit{0.00} & \textit{0.00} & \textit{0.00} & -8.59 & 5.85 & 1.47 & 1.44 (1.14, 1.76) \\ 
  WV & -2.05 & 3.08 & 0.67 & \textit{0.00} & \textit{0.00} & \textit{0.00} & -4.31 & 5.12 & 0.84 & 2.12 (1.32, 3.00) \\ 
  WY & \textit{0.00} & \textit{0.00} & \textit{0.00} & -1.02 & 0.91 & 1.13 & -11.12 & 3.19 & 3.49 & 2.40 (1.86, 2.93) \\ 
   \hline
\end{tabular}
\caption{Model comparison via expected log predictive density (ELPD). Shown are the difference in EPLD relative to the best model ($\Delta$ELPD), the standard error of ELPD difference (SE), and the number of standards error worse than the best model the ELPD difference is (\# SE). For each league, the best model is represented by an ELPD difference and associated standard errors of 0, which are italicised for reference. Using the best model in each league, we report the posterior mean and 95\% credible interval for home advantage during the 2023 season.}
\label{tab:elpd} 
\end{table}

\section{Discussion}\label{sec:discussion}

Home advantage in sports is a phenomenon whose existence is unequivocal yet whose drivers are poorly understood. Using a 20-year sample two-fold larger than any previously conducted in the literature (Table \ref{tab:lit_review}), we set out to understand temporal trends in the home advantage across all levels of American football in order to better understand its drivers. Our findings suggest that home advantage has declined significantly at the FBS level over the past twenty years, and perhaps to a lesser extent at other levels of college and professional football. This pattern, however, is not universal, and home advantage at the high school level remains largely unchanged (and perhaps slightly increasing in parts of the country) during that same time period. A natural question to ask is why does such heterogeneity exist, and why haven't the declines in home advantage at higher levels of American football trickled down to the high school level. Towards positing an answer to this question, it is necessary to understand what has changed across all levels of football in the past twenty years and what hasn't.

One possibility is use of replay and challenges to overturn incorrect calls, mitigating the impact of referee subjectivity by ensuring more eventual rulings are correct. In the NFL, coach's challenges have offered the ability for teams to request review of controversial plays since 1999, preceding the beginning of our sample in 2004 \cite{replay_review}. Since 2014, all challenges have been reviewed at a centralized league office in New York \cite{replay_review}. During the course of our sample, the NFL also added new rules to automatically review all scoring plays (beginning in 2011) and plays resulting in a turnover (beginning in 2012). Finally, beginning in 2022, the NFL now utilizes an expedited review system which allows offsite officials to quickly overrule on-field officials and overturn clear and obvious errors without the need for a coaches challenge or full review. Perhaps unsurprisingly, the rate of successful coaches challenges has nearly doubled over the course of our sample, from 31\% in 2004 to 58\% in 2022 \cite{replay_history_nfl}.

While the NFL has seen expanded use of replay review over the course of our sample, such improvements likely have less of an effect on changes in possible home advantage compared to the introduction of replay review to a league that didn't use it previously, as is the case in NCAA football. Experimental utilization of instant replay review in NCAA football began in 2004 in the Big 10 sample, and 2005 in other FBS conferences. It was not until 2006, however, that the NCAA Football Rules Committee officially approved use of replay review and published guidelines governing its use in games \cite{replay_review}. Nevertheless, this did not require teams to adopt the use of replay review at their home stadiums \cite{ncaa_rules} and smaller conferences did not initially adopt replay review \cite{replay_review}, owing to insufficient technology needed to run such a system. For analogous reasons, FCS schools were likely later to adopt widespread replay review compared to larger FBS schools, while even today, the majority of Division II/III games do not have access to sufficient angles in a timely matter necessary to run a replay review system. Finally, replay review was explicitly prohibited in high school football until 2019 \cite{replay_review}. Even today's its use is almost non-existent outside of perhaps state championship games played at large college or NFL stadiums. 

Another potential driver of home advantage differences between levels of American football is changes in travel. Undoubtedly, NFL teams haven't seen substantial changes in travel accommodations in the past two decades. While the jets teams fly on have probably gotten more luxurious, flying to games has been the norm for all years in our sample. Similarly, high school and low end college teams (e.g. Division III) also have not experienced has many changes in their modality of travel in the last two decades, with primary reliance on bus travel. On the other hand, top tier FBS teams have likely seen the biggest improvements towards how they travel between games compared to 20 years ago. While elite programs may have still flown between far games 20 years ago, plane travel is ubiquitous now, and the majority of FBS teams are flying chartered jets as opposed to commercial airfare. While plane travel exists in FCS and Division II, it's rarer \cite{flights}. If indeed implicit referee bias and travel are two possible drivers of home advantage, and FBS football has experienced the most substantial changes in both of these areas between 2004-2023, the sample considered in this paper, such reasons could potentially explain why FBS experienced the strongest decline in HA. 

Closely related to type of travel is distance traveled between games. Distance traveled is of interest because FBS teams will be required to travel significantly more miles in the coming years following conference realignments that broke traditional geographical ties. Unfortunately, including distance as a covariate using our data is difficult for two reasons. First, it might be easier for an FBS team to fly 500 miles to an away game than a Division III team to drive 250 miles to an away game. Furthermore, due to the lack granularity about location in our data, we only have longitude and latitude of each high school city, which yields a disproportionate number of games with incorrect estimates of 0 distance traveled. That is, there heterogeneity in the true distance of these labelled 0-distance games, and missing out on such information could lead to inaccurate modeling of the relationship between distance and home advantage.

\begin{figure}
    \centering
    \includegraphics[width = 0.8\textwidth]{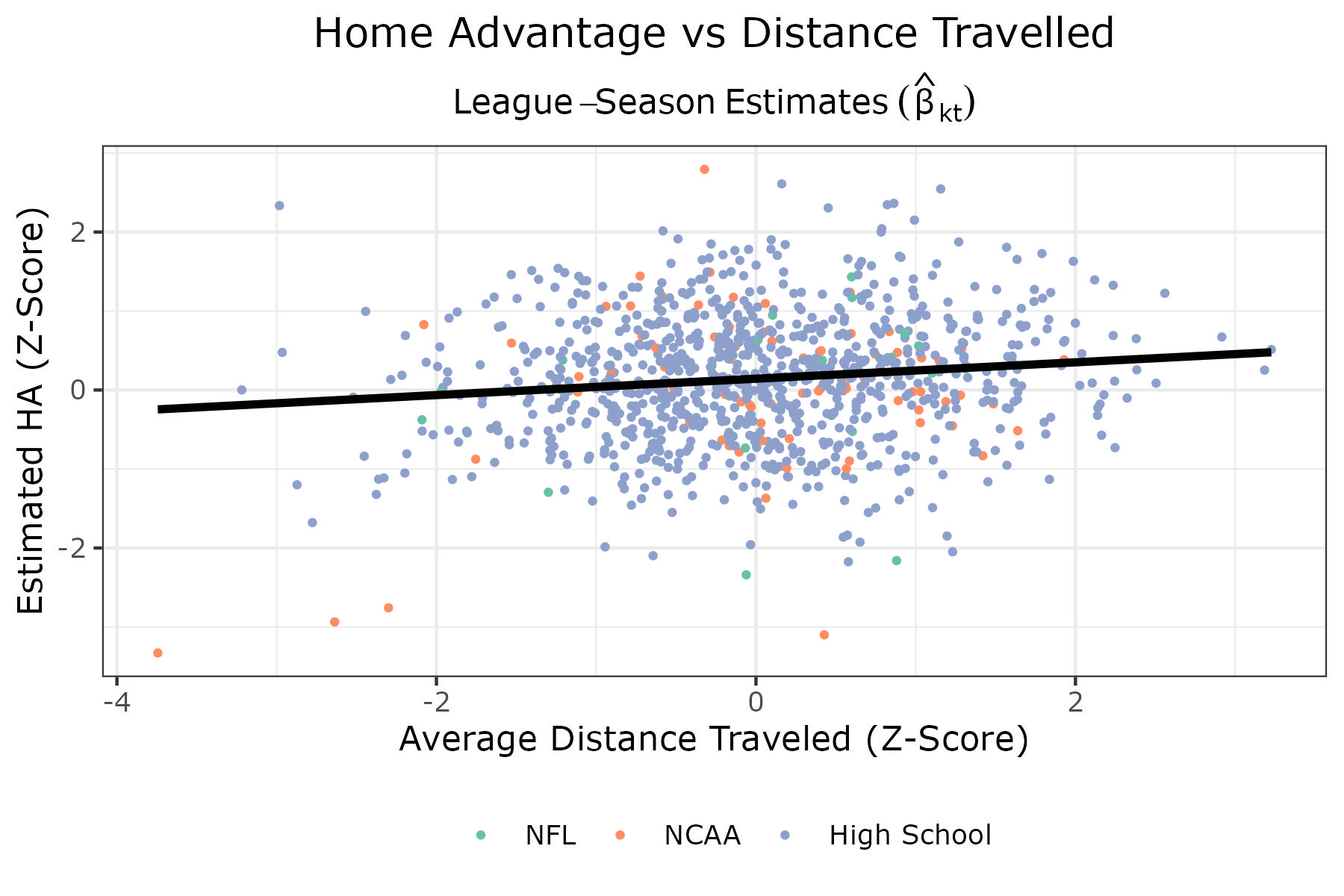}
    \caption{~Standardized mean distance travelled by the away team vs. standardized home advantage estimate $z^{(\gamma)}_{kt}$ across all league-seasons in the study. There is at best a weak association between average distanced travelled and estimated home advantage, with line of best fit different by roughly 0.5 SD in estimated home advantage between the extremes in average distance traveled. Overall, this suggests that changes in distance travelled between games is not the primary driver of the trends in home advantage over the past 20 years, at least not in the absence of also knowing method of travel.}
    \label{fig:alpha_distance}
\end{figure}

In order to understand the relationship between distance travelled and home advantage in our dataset, Figure \ref{fig:alpha_distance} displays standardized versions of $\hat \gamma_{kt}$ from Model \ref{eq:model3} against standardized average distance travelled in each league-season. Standardized home advantage estimates are defined as 

$$
z^{(\gamma)}_{kt} = \frac{\hat\gamma_{kt} - \bar \gamma}{\sigma_\gamma}
$$

\noindent where $\bar\gamma$ denotes the mean home advantage $\hat\gamma_{kt}$ across all leagues and seasons, and $\sigma_\gamma$ similarly denotes the standard deviation of home advantage posterior means across all leagues and seasons. To compute standardized distance travelled, we compute the average distance travelled by the away team in each league-season, and then compute the $z$-score of those quantities across all leagues/seasons in our sample. There is at best a weak association between average distanced travelled and estimated home advantage ($R^2 = 0.016$), suggesting that any change in distance between games is not the main factor driving these observed results. Adding further notion to the idea that distance travelled is perhaps secondary in driving changes in home advantage is the fact that no consistent pattern exists among rural states at the high school level. While Alaska, Montana, Wyoming, and Vermont had some of the largest estimated HA for the 2023 season, South Dakota, Idaho and Vermont's twin sister New Hampshire had some of the lowest home advantages in 2023.

While an even larger sample would be helpful to better test out some of these hypotheses further, doing so is not possible for lack of reliable high school scores. Furthermore, going back before 2004 would likely introduce additional confounding by play style, where changes in home advantage could also be attributed to rule changes that generally favor increases in offensive production. This is particularly important if one believes HA for football outcomes can model modeled on the multiplicative scale \cite{higgs2021}. Such changes also likely explain why works examining older seasons \cite{steffani1980, harville1980} have reported smaller HA estimates than more recent works \cite{glickman2017, david2011}, including our own work.

Comparison to other works in Table \ref{tab:lit_review} is imperfect because many papers use data outside the years of our sample and/or report home advantage on a win probability scale as opposed to a point differential scale.  Using our best NFL home advantage estimated of 1.73 points and associated temporal trend $\hat\beta_{1k} = -0.032$, we can obtain estimated of NFL home advantage in 2006 and 2014 of 2.27 points and 2.01 points, respectively, in line with the constant estimate of 2.4 points reported by \citet{glickman2017}. A similar computation for home advantage in FBS yields estimates of 3.65 and 3.85 during the 2008 and 2010 seasons respectively, slightly larger than the implied 3.1 points reported by \citet{wang2011}. These estimates are both a full point lower than the estimated NCAA home advantage during the 2013 and 2016 seasons reported by \citet{fullagar2019}, although that work did not exclude FBS vs. FCS games, which are nearly always hosted by superior FBS teams and generally result in a large margin of victory for the hosting team.

Our work also illustrates an intuitive but perhaps underappreciated point in the importance of properly adjusting for team strength when estimating home advantage.  In lower levels of football, including college and high school, better teams are both more likely to host more home games (perhaps due to better facilities and larger athletic budgets) and win by larger amounts. Additionally, in several leagues (including the NFL), better teams host playoff games, and so even with more balanced schedules, one would expect model-based estimates of HA which account for relative team strengths to be attenuated compared to empirical observations. As evidenced in Figure \ref{fig:ha_graphic}, failure to properly account for confounding by team strength would lead one to not only produce estimates of HA that are severely biased, but also overstate the extent to which strong temporal HA trends exist. Alternative methods that rely on empirical home win \% \cite{jones2016, pollard2015} or score differential \cite{mccutcheon2015, steffani1980} do not account for relative team strength, and thus provided an incomplete characterization of home advantage. Although the primary purpose of our paper was not to evaluate the ability to estimate team strength, we compared the team ratings $\theta_{ikt}$ obtained from the best model for each NCAA division to ratings from MasseyRatings.com \cite{masseyratings}, obtaining a correlation of 0.92, suggesting that methods utilized in this paper are able to estimate team strengths with well respected publicly available ratings. This is notable because estimation of home advantage is ultimately limited by the ability to model team strengths. Models that better estimate team strength will generally produce more accurate estimates of the home advantage \citep{benz2023estimating}.

Though a linear trend appears appropriate for the majority of leagues plotted in Figure \ref{fig:ha_graphic}, our modeling framework does not allow for the possibility of a non-linear relationship between time and changes in the home advantage. For example, low-dimensional polynomial fits (e.g. quadratic or cubic) or a spline term on time $t$ may be more appropriate for a single league of interest. These models would allow for more flexibility than the linear trend assumed by Model \ref{eq:model2}, while also requiring fewer parameters than Model \ref{eq:model3}. So while our framework is useful for assessing large-scale temporal trends across leagues, it is plausible one of these other candidate fits would yield improved inference. We leave consideration of additional candidate models as an area of possible exploration for future work.

Another possible limitation of our framework is that no information is shared from state to state or league to league, as we fit the same model structure separately. In certain contexts, however, it may be beneficial to share information across leagues, particularly if there is some justifiable reason why they might be related. To explore this consideration further, we investigated a hierarchical version of Model 
 \ref{eq:model2}  applied to high school states. Two worked examples, along with comparison to results of Model \ref{eq:model2} and a discussion of the hierarchical framing, are presented in the Supplementary Materials.

 Finally, another possible extension of this work would consider various correlation structures on $\gamma_{kt}$ in Model \ref{eq:model3}, possibly through application of a dynamic state space model \citep{glickman1998, glickman2017, lopez2018}. For example, understanding the autocorrelation between $\gamma_{kt}$ and $\gamma_{k(t-1)}$ could be of interest, or groupings of $\gamma_{kt}$ across leagues over time by factors such as replay usage or rule changes. One of the benefits of Model \ref{eq:model3} is it's ability to capture fast changes in HA in response to massive systematic changes, such as the introduction of replay or return to play following COVID. While effects of this nature are not the primary focus of this work, improved exploration of these correlation structures across both leagues and time could yield improved insight.

Overall, our work shows that conclusions from professional leagues need not generalize to amateur leagues, and provides a simple yet robust framework for creation and evaluation of models to understand the home advantage in American football which scales well with large numbers of teams and high rates of player turnover. This work is the first to consider a comprehensive evaluation of the home advantage of American football across levels of the sport, and does so at a scale far larger than any previous study to date, particularly at the high school level. Home advantage is notoriously difficult to study, and is likely the combination of numerous interdependent factors. In the absence of rare events (e.g. Covid-19 limiting fans in stands \cite{higgs2021}), it's often impossible to isolate and study any single factor contributing to home advantage. By considering all levels of American football, not only do we get a more complete understanding of the HA landscape, but also we can better hypothesize which factors are driving home advantage by comparing what is different across levels of the game.

\section*{Data and Code Availability}
All data and code used for this analysis is publicly available on GitHub at \url{https://github.com/ThompsonJamesBliss/comprehensive_survey_american_football_home_adv} .

\clearpage


\begin{thebibliography}{48}
\providecommand{\natexlab}[1]{#1}
\providecommand{\url}[1]{\texttt{#1}}
\expandafter\ifx\csname urlstyle\endcsname\relax
  \providecommand{\doi}[1]{doi: #1}\else
  \providecommand{\doi}{doi: \begingroup \urlstyle{rm}\Url}\fi

\bibitem[King(2024)]{kingFMIA}
Peter King.
\newblock {FMIA Wild Card: Decibels in Detroit, Doubt in Dallas, Harmony in
  Houston}.
\newblock Football Morning in America, 2024.
\newblock URL
  \url{https://www.nbcsports.com/nfl/profootballtalk/fmia/news/nfl-wild-card-playoffs-detroit-lions-dan-campbell-cj-stroud-peter-king-fmia}.

\bibitem[Haberstroh(2015)]{haberstroh2015}
Tom Haberstroh.
\newblock Home-court advantage? not so much, 2015.
\newblock URL
  \url{https://www.espn.com/nba/story/_/id/12241619/home-court-advantage-decline}.

\bibitem[Haberstroh and Illardi(2015)]{haberstroh2015b}
Tom Haberstroh and Steve Illardi.
\newblock Nba home court losing its value, 2015.
\newblock URL
  \url{https://www.espn.com/nba/insider/story/_/id/12243076/nba-analyzing-diminishing-value-home-court-advantage}.

\bibitem[Kent(2017)]{kent2017}
Brendan Kent.
\newblock {NBA Home-Count Advantage is in Decline, are 3s to Blame?}, 2017.
\newblock URL
  \url{https://harvardsportsanalysis.org/2017/03/nba-home-court-advantage-is-in-decline-are-3s-to-blame/}.

\bibitem[Davis(2017)]{davis2017}
Seth Davis.
\newblock Hoop thoughts: What is happening to home court advantage?
\newblock Sports Illustrated, 2017.
\newblock URL
  \url{https://www.si.com/college/2017/01/09/hoop-thoughts-home-court-advantage-baylor-kansas-villanova}.

\bibitem[Pomeroy(2017)]{kenpom2017}
Ken Pomeroy.
\newblock How to measure site-specific home-court advantage, part two.
\newblock KenPom, 2017.
\newblock URL
  \url{https://kenpom.com/blog/how-to-measure-site-specific-home-court-advantage-part-two/}.

\bibitem[Jones(2019)]{jones2019}
M.~B. Jones.
\newblock The sustained reduction-by-half of home advantage in the nhl,
  1991-1992 to 2000-2001.
\newblock \emph{SAGE Open}, 9\penalty0 (1), 2019.
\newblock \doi{10.1177/2158244019830008}.
\newblock URL \url{https://doi.org/10.1177/2158244019830008}.

\bibitem[Lopez(2016)]{lopesz2016}
Michael Lopez.
\newblock On soccer's declining home field advantage, 2016.
\newblock URL
  \url{https://statsbylopez.com/2016/05/13/on-soccers-declining-home-field-advantage/}.

\bibitem[Ooeder and Curley(2014)]{roeder2014}
Oliver Ooeder and James Curley.
\newblock {Home-Field Advantage Doesn’t Mean What It Used To In English
  Football}.
\newblock FiveThirtyEight, 2014.
\newblock URL
  \url{https://fivethirtyeight.com/features/home-field-advantage-english-premier-league/}.

\bibitem[{Stan Development Team}(2019)]{standocs}
{Stan Development Team}.
\newblock Stan reference manual, 2019.
\newblock URL \url{https://mc-stan.org/docs/2_26/reference-manual/index.html}.

\bibitem[Vehtari et~al.(2017)Vehtari, Gelman, and Gabry]{vehtari2017practical}
Aki Vehtari, Andrew Gelman, and Jonah Gabry.
\newblock Practical bayesian model evaluation using leave-one-out
  cross-validation and waic.
\newblock \emph{Statistical Computing}, 27\penalty0 (6):\penalty0 1413--1432,
  2017.
\newblock \doi{10.1007/s11222-016-9696-4}.
\newblock URL \url{https://doi.org/10.1007/s11222-016-9696-4}.

\bibitem[Higgs and Stavness(2021)]{higgs2021}
Nico Higgs and Ian Stavness.
\newblock {Bayesian analysis of home advantage in North American professional
  sports before and during COVID-19}.
\newblock \emph{Nature Scientific Reports}, 11, 2021.

\bibitem[Lopez et~al.(2018)Lopez, Matthews, Baumer, et~al.]{lopez2018}
Michael~J Lopez, Gregory~J Matthews, Benjamin~S Baumer, et~al.
\newblock {How often does the best team win? A unified approach to
  understanding randomness in North American sport}.
\newblock \emph{The Annals of Applied Statistics}, 12\penalty0 (4):\penalty0
  2483--2516, 2018.

\bibitem[Glickman and Stern(2017)]{glickman2017}
Mark Glickman and Hal Stern.
\newblock {Estimating team strength in the NFL}.
\newblock In Jim Albert, Mark~E. Glickman, Tim~B. Swartz, and Ruud~H. Koning,
  editors, \emph{Handbook of Statistical Methods and Analyses in Sports},
  chapter~5, pages 113--136. CRC Press, Boca Raton, FL, 2017.

\bibitem[Jones(2016)]{jones2016}
Marshall~B. Jones.
\newblock {Injuries and home advantage in the NFL}.
\newblock \emph{SpringerPlus}, 5, 2016.

\bibitem[David et~al.(2011)David, Pasteur, Ahmad, and Janning]{david2011}
John~A. David, R.~Drew Pasteur, M.~Saif Ahmad, and Michael~C. Janning.
\newblock {NFL Prediction using Committees of Artificial Neural Networks}.
\newblock \emph{Journal of Quantitative Analysis in Sports}, 7\penalty0 (2),
  2011.

\bibitem[Pollard and Gonzalez(2015)]{pollard2015}
Richard Pollard and Miguel Gonzalez.
\newblock {Comparison of Home Advantage in College and Professional Team Sports
  in the United States}.
\newblock \emph{Collegium Antropologicum}, 39\penalty0 (3):\penalty0 583--589,
  2015.

\bibitem[Baker and McHale(2013)]{baker2013}
Rose Baker and Ian McHale.
\newblock {Forecasting exact scores in National Football League games}.
\newblock \emph{International Journal of Forecasting}, 29\penalty0
  (1):\penalty0 122--130, 2013.

\bibitem[Glickman and Stern(1998)]{glickman1998}
Mark Glickman and Hal Stern.
\newblock {A State-Space Model for National Football League Scores}.
\newblock \emph{Journal of the American Statistical Association}, 93\penalty0
  (441):\penalty0 25--35, 1998.

\bibitem[Boulier and Stekler(2003)]{boulier2003}
Bryan Boulier and H.O. Stekler.
\newblock {Predicting the outcomes of National League Football games}.
\newblock \emph{International Journal of Forecasting}, 19:\penalty0 257--270,
  2003.

\bibitem[Steffani(1980)]{steffani1980}
Raymond Steffani.
\newblock {Improved Least Squares Football, Basketball, and Soccer
  Predictions}.
\newblock \emph{IEEE Transactions on Systems, Man, and Cybernetics},
  10:\penalty0 116--123, 1980.

\bibitem[Harville(1980)]{harville1980}
David Harville.
\newblock {Predictions for National Football League Games Via Linear-Model
  Methodology}.
\newblock \emph{Journal of the American Statistical Association}, 75\penalty0
  (361):\penalty0 516--524, 1980.

\bibitem[Fullagar et~al.(2019)Fullagar, Delaney, Duffield, and
  Murray]{fullagar2019}
Hugh H.~K. Fullagar, Jace Delaney, Rob Duffield, and Andrew Murray.
\newblock {Factors influencing home advantage in American collegiate football}.
\newblock \emph{Science and Medicine in Football}, 3\penalty0 (2):\penalty0
  163--168, 2019.
\newblock \doi{10.1080/24733938.2018.1524581}.

\bibitem[Wang et~al.(2011)Wang, Johnston, and Jones]{wang2011}
Winnie Wang, Ron Johnston, and Kelvyn Jones.
\newblock {Home Advantage in American College Football Games: A Multilevel
  Modeling Approach}.
\newblock \emph{Journal of Quantitative Analysis in Sports}, 7, 2011.

\bibitem[Caudill and Mixon(2007)]{caudill2007}
Steven Caudill and Franklin Mixon.
\newblock {Stadium size, ticket allotments, and home field advantage in college
  football}.
\newblock \emph{The Social Science Journal}, 44:\penalty0 751--759, 2007.

\bibitem[Gajewski(2006)]{gajewski2006}
Byron~J Gajewski.
\newblock {There's No Place Like Home: Estimating Intra-Conference Home Field
  Advantage in College Football Using a Bayesian Piecewise Linear Model}.
\newblock \emph{Journal of Quantitative Analysis in Sports}, 2\penalty0 (1),
  2006.
\newblock \doi{doi:10.2202/1559-0410.1009}.

\bibitem[Massey(1997)]{massey1997}
Kenneth Massey.
\newblock {Statistical Models Applied to the Rating of Sports Teams}.
\newblock Bluefield College Honors Thesis, 1997.
\newblock URL \url{https://masseyratings.com/theory/massey97.pdf}.

\bibitem[Harville(1977)]{harville1977}
David Harville.
\newblock {The Use of Linear-Model Methodology to Rate High School or College
  Football Teams}.
\newblock \emph{Journal of the American Statistical Association}, 72\penalty0
  (358):\penalty0 278--289, 1977.

\bibitem[McCutcheon(1984)]{mccutcheon2015}
Lynn~E McCutcheon.
\newblock {The Home Advantage in High School Athletics}.
\newblock \emph{Journal of Sport Behavior}, 7\penalty0 (4), 1984.

\bibitem[Schwartz and Barsky(1977)]{schwartz1977home}
Barry Schwartz and Stephen~F Barsky.
\newblock The home advantage.
\newblock \emph{Social forces}, 55\penalty0 (3):\penalty0 641--661, 1977.

\bibitem[Moskowitz and Wertheim(2011)]{moskowitz2011scorecasting}
Tobias Moskowitz and L~Jon Wertheim.
\newblock \emph{Scorecasting: The hidden influences behind how sports are
  played and games are won}.
\newblock Crown Archetype, 2011.

\bibitem[Snyder and Lopez(2015)]{snyder2015consistency}
Kevin Snyder and Michael Lopez.
\newblock Consistency, accuracy, and fairness: a study of discretionary
  penalties in the nfl.
\newblock \emph{Journal of Quantitative Analysis in Sports}, 11\penalty0
  (4):\penalty0 219--230, 2015.

\bibitem[Vergin and Sosik(1999)]{vergin1999no}
Roger~C Vergin and John~J Sosik.
\newblock No place like home: an examination of the home field advantage in
  gambling strategies in nfl football.
\newblock \emph{Journal of Economics and Business}, 51\penalty0 (1):\penalty0
  21--31, 1999.

\bibitem[Nichols(2014)]{nichols2014impact}
Mark~W Nichols.
\newblock The impact of visiting team travel on game outcome and biases in nfl
  betting markets.
\newblock \emph{Journal of Sports Economics}, 15\penalty0 (1):\penalty0 78--96,
  2014.

\bibitem[Benz and Lopez(2023)]{benz2023estimating}
Luke~S Benz and Michael~J Lopez.
\newblock Estimating the change in soccer’s home advantage during the
  covid-19 pandemic using bivariate poisson regression.
\newblock \emph{AStA Advances in Statistical Analysis}, 107\penalty0
  (1-2):\penalty0 205--232, 2023.

\bibitem[Gelman and Rubin(1992)]{gelman1992}
Andrew Gelman and Donald~B. Rubin.
\newblock Inference from iterative simulation using multiple sequences.
\newblock \emph{Statist. Sci.}, 7\penalty0 (4):\penalty0 457--472, 1992.

\bibitem[Brooks and Gelman(1998)]{brooks1998}
Stephen~P. Brooks and Andrew Gelman.
\newblock General methods for monitoring convergence of iterative simulations.
\newblock \emph{Journal of Computational and Graphical Statistics}, 7\penalty0
  (4):\penalty0 434--455, 1998.

\bibitem[Gelman et~al.(2013)Gelman, Carlin, Stern, Dunson, Vehtari, and
  Rubin]{bda3}
Andrew Gelman, John~B. Carlin, Hal~S. Stern, David~B. Dunson, Aki Vehtari, and
  Donald~B Rubin.
\newblock \emph{Bayesian Data Analysis}.
\newblock CRC Press, Boca Raton, FL, 3 edition, 2013.
\newblock ISBN 9781439840955.

\bibitem[Vehtari et~al.(2023)Vehtari, Gabry, Magnusson, Yao, Bürkner,
  Paananen, and Gelman]{loopkg}
Aki Vehtari, Jonah Gabry, Mans Magnusson, Yuling Yao, Paul-Christian Bürkner,
  Topi Paananen, and Andrew Gelman.
\newblock loo: Efficient leave-one-out cross-validation and waic for bayesian
  models, 2023.
\newblock URL \url{https://mc-stan.org/loo/}.
\newblock R package version 2.6.0.

\bibitem[Carl and Baldwin(2023)]{nflfastr}
Sebastian Carl and Ben Baldwin.
\newblock \emph{nflfastR: Functions to Efficiently Access NFL Play by Play
  Data}, 2023.
\newblock https://www.nflfastr.com/, https://github.com/nflverse/nflfastR.

\bibitem[Massey(2023)]{masseyratings}
Kenneth Massey, 2023.
\newblock URL \url{https://masseyratings.com/ranks?s=cf}.

\bibitem[max(2023)]{maxpreps}
High school football, 2023.
\newblock URL \url{https://www.maxpreps.com/football/}.

\bibitem[Vehtari(2020)]{vehtari2020}
Aki Vehtari.
\newblock More limitations of cross-validation and actionable recommendations,
  2020.
\newblock URL
  \url{https://statmodeling.stat.columbia.edu/2020/08/27/more-limitations-of-cross-validation-and-actionable-recommendations/}.

\bibitem[League(2023)]{txrules}
Texas University~Interscholastic League.
\newblock \emph{Football Manual Post Season Information}, 2023.
\newblock
  https://www.uiltexas.org/football/manual/football-manual-post-season-information.

\bibitem[{Wikipedia contributors}(2023)]{replay_review}
{Wikipedia contributors}.
\newblock Replay review in gridiron football, 2023.
\newblock URL
  \url{https://en.wikipedia.org/wiki/Replay_review_in_gridiron_football}.

\bibitem[{NFL Football Operations}(2023)]{replay_history_nfl}
{NFL Football Operations}.
\newblock History of instant replay, 2023.
\newblock URL
  \url{https://operations.nfl.com/officiating/instant-replay/history-of-instant-replay/}.

\bibitem[{NCAA}(2023)]{ncaa_rules}
{NCAA}.
\newblock \emph{2023 NCAA Football Instant Replay Case Book}.
\newblock 2023.

\bibitem[McCready and Kroll()]{flights}
Bo~McCready and Doug Kroll.
\newblock An examination of college football flight patterns.
\newblock URL
  \url{https://athleticdirectoru.com/articles/an-examination-of-ncaa-college-football-flight-patterns-2/}.

\end{thebibliography}
\end{document}


\articletype{Supplementary Materials}


\title{A comprehensive survey of the home advantage in American football}
\runningtitle{Supplementary Materials: Home Advantage in American Football}


\author[1]{Luke Benz$^\dagger$}
\author[2]{Thompson Bliss$^\dagger$}
\author[3]{Michael Lopez} 
\runningauthor{Benz, Bliss, and Lopez}
\affil[1]{\protect\raggedright 
Department of Biostatistics, Harvard T.H. Chan School of Public Health, Boston MA  (lukebenz@g.harvard.edu)}
\affil[2]{\protect\raggedright 
National Football League, New York, NY (thompson.bliss@nfl.com)}
\affil[3]{\protect\raggedright 
National Football League, New York, NY (michael.lopez@nfl.com)}


\def\thefootnote{$^\dagger$}\footnotetext{These authors contributed equally to this work}\def\thefootnote{\arabic{footnote}}

\maketitle

\section{Hierarchical Modeling Approach}\label{sec:s1}
An alternative way to fit Model 2 involves a Bayesian hierarchical model, which has been used in several sports related problems in recent years \cite{baio, gabrio2020bayesian, hobbs2020,ingram2019point, robinson2020}. In the context of our problem, this approach may be appealing when estimating league specific home advantage trends because it shares information across leagues. This assumes that $\beta_{1k} \sim N(\beta_1^*, \lambda_1^2)$ where $\beta_1^*$ denotes some common trend. Under this framework, $\beta_{1}^*$ uses information across all leagues and $\beta_{1k}$ is shrunk towards $\beta_{1}^*$. Implicitly, these assumptions imply a shared structure between leagues. 

In terms of the 4 levels of the NCAA football, we feel that these divisions are sufficiently different from one another that such a model is not justified. That is, differences between Division III (most away teams travel via bus, few NFL players) and FBS (most away teams travel via plane, several NFL players) are stark enough in both travel and team ability, a finding supported by Table 3 in the main body of our paper. 

We feel instead that high school football presents a better motivating example to illustrate the use of this model. In Section \ref{sec:s1.1} present a mathematical formulation of the hierarchical version Model 2, which we refer to as Model 2H. We present two examples of Model 2H, one on states contributing fewer than 10,000 games to our collective sample (Section \ref{sec:s1.2}) and one examining all 50 states together (Section \ref{sec:s1.3}). Finally, in Section \ref{sec:s1.4}, we offer some discussion on the two examples and some commentary on when a model like Model 2H is well justified.

\subsection{Model Formulation}\label{sec:s1.1}

As in the main paper, let $Y_{ijkt}$ be the score differential in a game between team $i$ and team $j$ in league $k$ during year $t$. We assume that 
$$
Y_{ijkt} \sim N(\mu_{ijkt}, \sigma^2_{k})
$$

\noindent Just as in Model 2, we assume

$$
\mu_{ijkt} = \theta_{ikt} - \theta_{jkt} + \beta_{0k} + \beta_{1k}(t - t_0)
$$

\noindent Where Model 2 and Model 2H differ is their priors. Specifically, for Model 2H, we assume the following.

$$
\begin{aligned}
  \theta_{ikt} &\sim N(0, \zeta^2_{k})~~~\text{(Team Strengths)} \\
  \zeta_k &\sim \text{HalfNormal}(0,5^2)~~~\text{(Team Strength Variance)}  \\
  \sigma_k  &\sim \text{HalfNormal}(0,5^2) ~~~\text{(Score Differential Variance)} \\ 
  \beta_{0k} &\sim \text{Normal}(0, \lambda_{0k}^2) ~~~\text{(League Specific Home Advantage Intercept)} \\
  \beta_{1k} &\sim \text{Normal}(\alpha_1^*, \lambda_{1}^2) ~~~\text{(League Specific Home Advantage Trend)} \\
  \beta_{1}^* &\sim \text{Normal}(0, 5^2) ~~~\text{(Shared Home Advantage Trend)} \\
  \lambda_{0k} &\sim \text{HalfNormal}(0,5^2) ~~~\text{(League Specific  Home Advantage Intercept Variance)} \\
  \lambda_{1k} &\sim \text{HalfNormal}(0,5^2) ~~~\text{(League Specific Home Advantage Trend Variance)}  \\ \\
  \nonumber
\end{aligned}
$$

While we put shared structure on state specific home advantage trends $\beta_{1k}$, we choose not to assume a shared structure on intercepts $\beta_{0k}$ because data begins in different years for different states and thus the reference season indexed by the intercept in each states differs. For example, $\beta_{0,\text{Texas}}$ indexes home advantage in Texas in 2004 while $\beta_{0, \text{Oregon}}$ indexes home advantage in Oregon in 2007. As with our other models, models in each example below were fit using 4 parallel chains, each made up of 2000 iterations, and a burn in of 500 draws.

\subsection{Example 1: Small States}\label{sec:s1.2}

\begin{figure}[H]
    \centering
    \includegraphics[width=\textwidth]{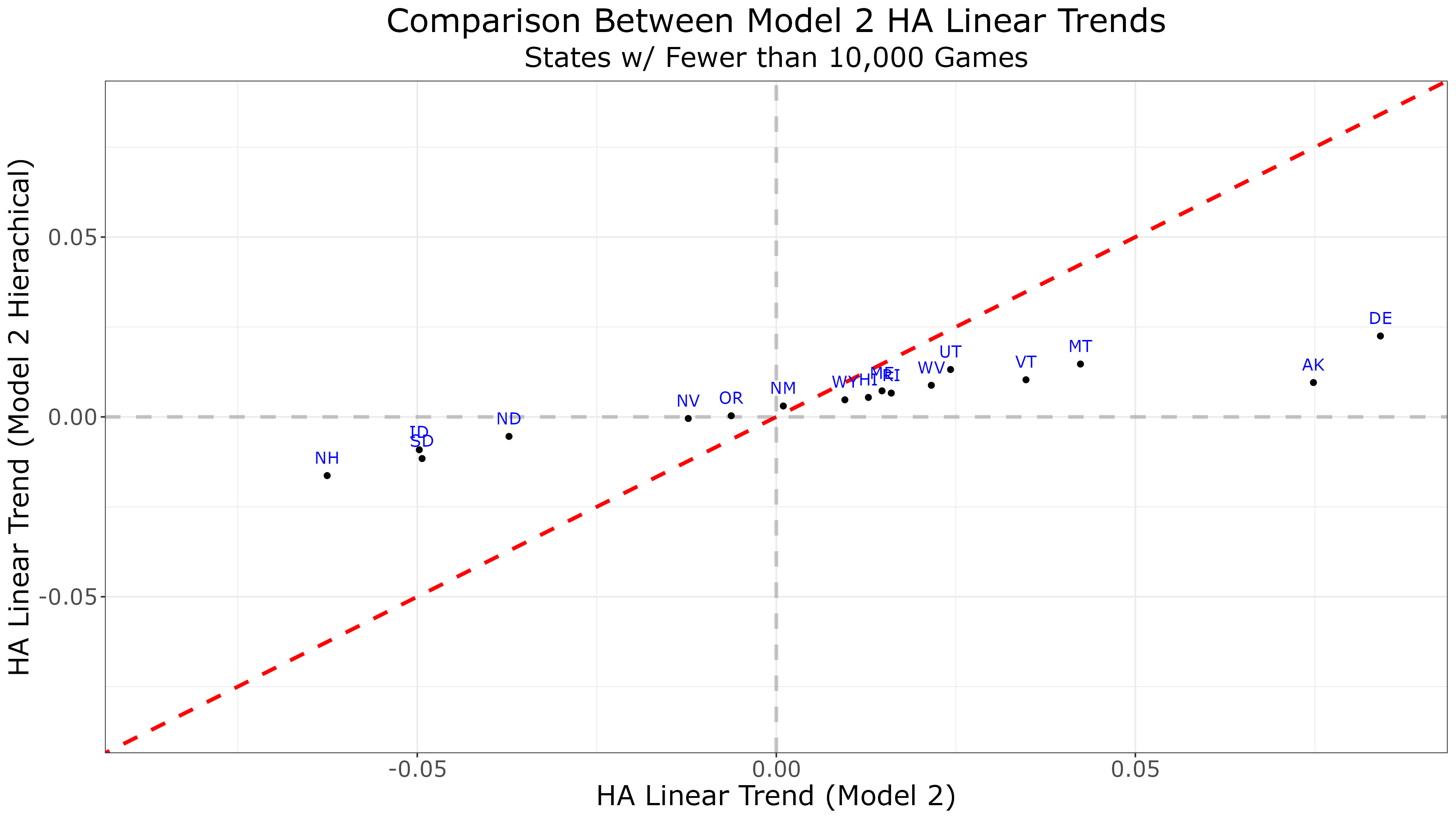}
    \caption{Comparison of posterior means $\alpha_{1k}$ from Model 2 and Model 2H when Model 2H was fit on states with fewer than 10,000 games}.
    \label{fig:small_trend_comp}
\end{figure}
We fit Model 2H on the 17 states which contributed fewer than 10,000 games to our overall sample (Table \ref{tab:sample_size_state}). Though these states are not geographically close, they are all generally rural and not population dense. Figure \ref{fig:small_trend_comp} compares posterior means $\hat \beta_{1k}$ under the original version of Model 2 and the hierarchical version of Model 2. Unsurprisingly, we observe shrinkage towards $\hat \beta_{1}^* = 0.004$, and estimate the associated standard deviation around this shared trend to be $\hat \lambda_1 = 0.022$. With the exception of Alaska, the state with the smallest sample size, the order of $\hat \beta_{1k}$ is roughly preserved with New Hampshire and Delaware on the respective extremes. 

\begin{figure}[H]
    \centering
    \includegraphics[width=\textwidth]{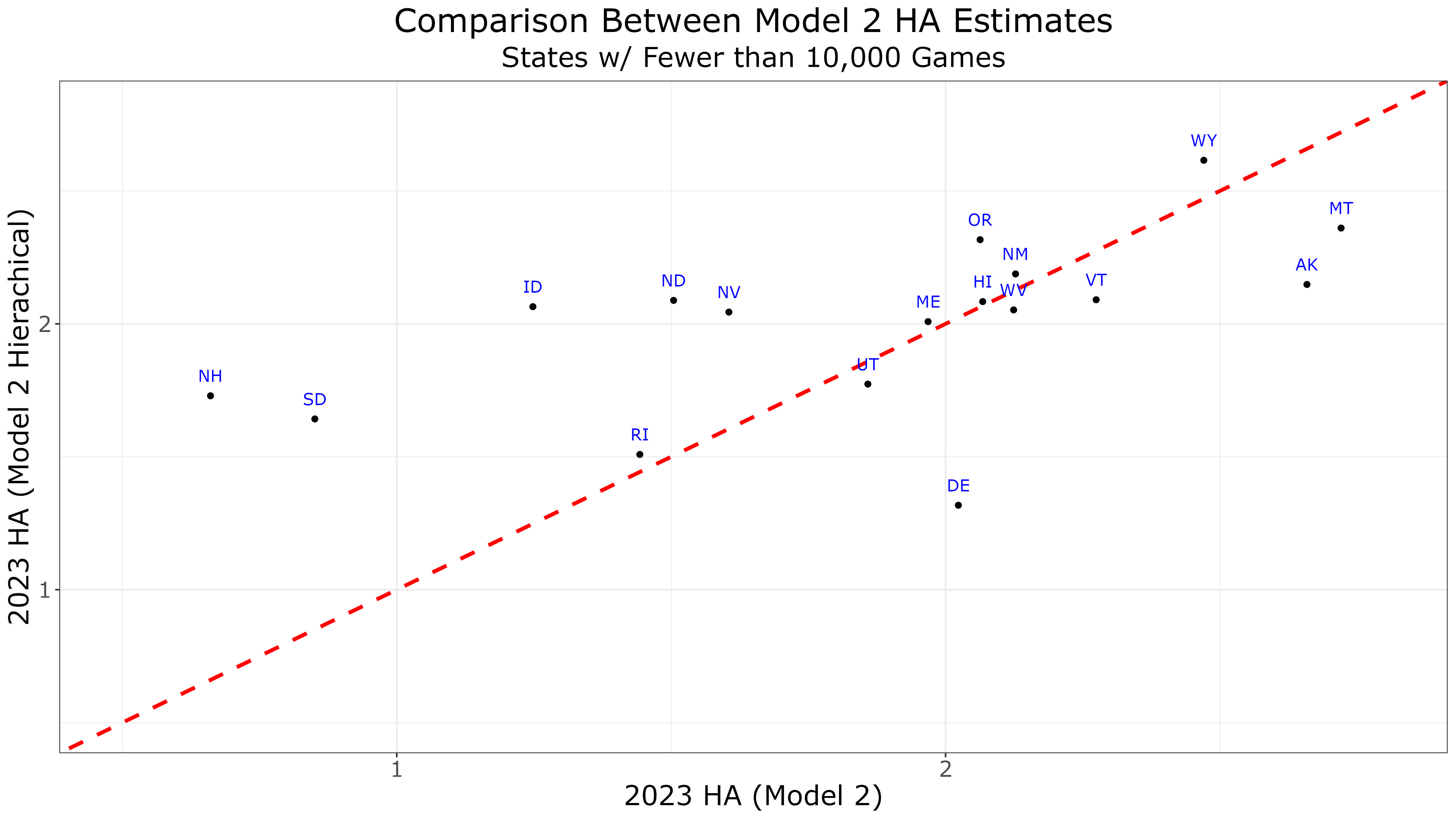}
    \caption{Comparison of posterior means for home advantage from Model 2 and Model 2H when Model 2H was fit on states with fewer than 10,000 games}.
    \label{fig:small_HA_comp}
\end{figure}

Significance of certain state specific trends are attenuated compared to fitting each league in isolation. For example, under Model 2, $P(\hat\beta_{1k} > 0) = 0.968$ for Delaware, while under Model 2H, $P(\hat\beta_{1k} > 0) = 0.808$. Though the magnitudes of state specific HA trends may be attenuated, broader conclusions remain the same. Namely, there is not any evidence to suggest HA is in decline in high school school (as is likely the case in higher leagues) and there is a degree of heterogeneity in both home advantage (Figure \ref{fig:small_HA_comp}) and home advantage trends at the high school level.

\subsection{Example 2: All 50 States}\label{sec:s1.3}

\begin{figure}[H]
    \centering
    \includegraphics[width=\textwidth]{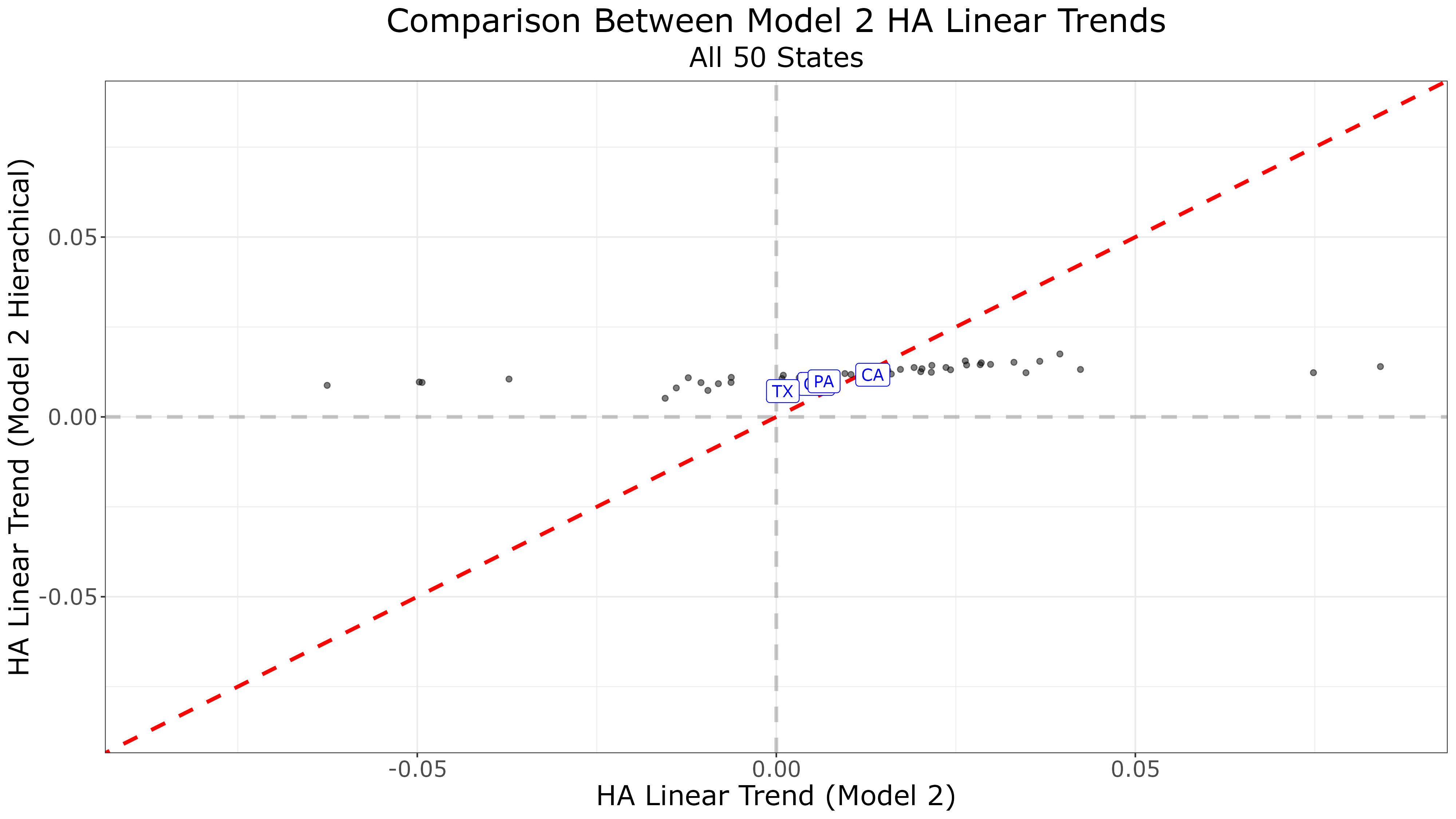}
    \caption{Comparison of posterior means $\beta_{1k}$ from Model 2 and Model 2H when Model 2H was fit on all 50 states. States which contributed the four largest sample sizes are highlighted in blue, and do not exhibit much shrinkage.}
    \label{fig:trend_comp}
\end{figure}

We also fit Model 2H on all 50 states together. Figure \ref{fig:trend_comp} compares posterior means $\hat \beta_{1k}$ under the original version of Model 2 and the hierarchical version of Model 2. State effects are shrunk substantially closer to $\hat\beta_{1^*} = 0.012$ than in the previous example. Furthermore, the posterior mean standard deviation of state trends $\hat \lambda_{1} =  0.006$ seems almost implausibly small. In Figure \ref{fig:trend_comp}, Texas, California, Pennsylvania, and Ohio, the four states which contributed the largest sample of games (Table \ref{tab:sample_size_state}) are highlighted in blue. Notably, these states do not exhibit much, if any, shrinkage. Collectively, these states contributed roughly 27\% of all high school games. 

\begin{figure}[H]
    \centering
    \includegraphics[width=\textwidth]{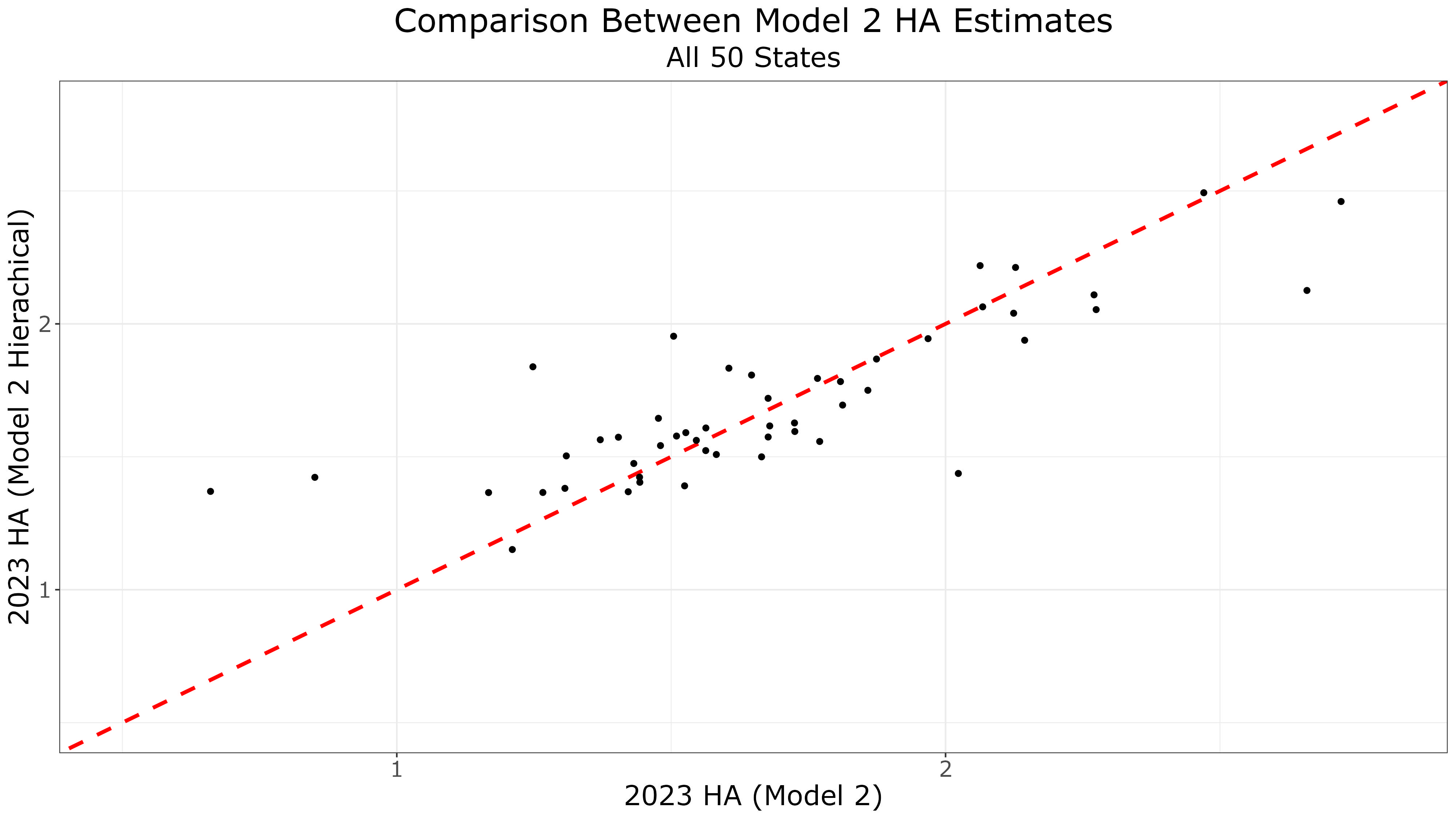}
    \caption{Comparison of posterior means for home advantage from Model 2 and Model 2H when Model 2H was fit on all 50 states}.
    \label{fig:HA_comp}
\end{figure}

Overall, when fit on all 50 states, Model 2H seems to exhibit over shrinkage. Reasonably, we'd expected $\beta_{1}^*$ to be closer to 0 as in the previous example with $\lambda_1$ larger that which was estimated in this example. Despite the fact that state specific linear trends $\hat \beta_{1k}$ are shrunk considerably towards $0.012$, 2023 HA estimates using Model 2H are not substantially different than those using Model 2 (Figure \ref{fig:HA_comp}), suggesting that state specific intercepts $\beta_{0k}$ are correcting for some of the observed over shrinkage.

\begin{table}[H]
\centering
\begin{tabular}{cc}
  \hline
League & \# of Games \\ 
  \hline
TX & 115,261 \\ 
  CA & 105,208 \\ 
  OH & 67,684 \\ 
  PA & 56,720 \\ 
  FL & 50,519 \\ 
  NY & 46,785 \\ 
  IL & 46,708 \\ 
  GA & 45,343 \\ 
  MI & 44,894 \\ 
  AL & 42,855 \\ 
  NC & 41,586 \\ 
  WI & 35,799 \\ 
  TN & 33,813 \\ 
  NJ & 32,482 \\ 
  IN & 31,242 \\ 
  VA & 30,233 \\ 
  MO & 29,894 \\ 
  MA & 27,751 \\ 
  LA & 27,672 \\ 
  OK & 26,403 \\ 
  MS & 25,885 \\ 
  MN & 25,439 \\ 
  IA & 24,170 \\ 
  WA & 23,272 \\ 
  SC & 23,239 \\ 
  \hline
  \end{tabular}
  \begin{tabular}{cc}
  \hline
League & \# of Games \\ 
  \hline
  
  KY & 22,701 \\ 
  AR & 20,487 \\ 
  KS & 20,093 \\ 
  AZ & 19,504 \\ 
  CO & 19,470 \\ 
  CT & 13,997 \\ 
  MD & 13,429 \\ 
  NE & 12,765 \\ 
  UT & 9,946 \\ 
  OR & 8,942 \\ 
  NM & 6,471 \\ 
  WV & 5,320 \\ 
  ME & 5,270 \\ 
  MT & 4,876 \\ 
  NH & 4,826 \\ 
  NV & 4,761 \\ 
  SD & 4,596 \\ 
  RI & 4,389 \\ 
  ND & 4,158 \\ 
  HI & 3,463 \\ 
  ID & 3,430 \\ 
  DE & 3,122 \\ 
  VT & 2,902 \\ 
  WY & 2,663 \\ 
  AK & 1,093 \\ 
  \hline
\end{tabular}
\caption{\# of games in each high school state.}\label{tab:sample_size_state}
\end{table}

\subsection{Discussion}\label{sec:s1.4}

The two examples in the preceding sections are illustrative of the differences between modeling approaches which fit a separate model for each league, as we did in the main body of our paper, and a hierarchical approach. These new examples suggest that a hierarchical approach, at least in our specification, may yield over shrinkage for estimating trends in home advantage when leagues are extremely different in size. 

Estimates of the 2023 home advantage seem to be less affected by over shrinkage than do trends $\beta_{1k}$. New Hampshire (Model 2: 0.66 , Model 2H: 1.37) and South Dakota (Model 2: 0.85 , Model 2H: 1.42), for example, are both reasonably pulled towards the overall high school average. One possibility is that because intercepts $\beta_{0k}$ are not estimated with a shared intercept, state-level HA estimates are not pulled towards a common, high school level HA. In fact, state-level intercept estimates may be reacting to the degree of shrinkage in state trends. Additional hierarchical models with both shared intercepts and shared linear trends, or models that incorporate spatial and/or rules-based similarities between states, are left as future work.

\section{Model Diagnostics}

Model diagnostics to assess convergence in the form of $\hat R$ statistics and effective sample sizes are presented for Models 1-3 in Tables \ref{tab:s1} - \ref{tab:s3}  respectively. Similar diagnostic information for the hierarchical models discussed in the previous section is available in Table \ref{tab:s4}. All $\hat R$ were very close to 1 indicating good model convergence \cite{gelman1992, brooks1998}.

\begin{table}[ht]
\centering
\begin{tabular}{cccccccc}
  \hline
League & $\min(\hat R)$ & $\max(\hat R)$ & $\alpha_k$ & $\eta_k$ & $\sigma_k$ & $\zeta_k$ & $\theta_k$  \\ 
  \hline
AK & 1.00 & 1.00 & 7639 & 6281 & 4723 & 5689 & 5553 \\ 
  AL & 1.00 & 1.00 & 8757 & 7550 & 4305 & 4010 & 5090 \\ 
  AR & 1.00 & 1.01 & 6772 & 5798 & 4138 & 1627 & 2602 \\ 
  AZ & 1.00 & 1.02 & 5404 & 5079 & 3899 & 1275 & 1245 \\ 
  CA & 1.00 & 1.01 & 8971 & 6996 & 3441 & 2456 & 2905 \\ 
  CO & 1.00 & 1.01 & 5798 & 5538 & 3686 & 1587 & 1694 \\ 
  CT & 1.00 & 1.01 & 4382 & 4866 & 4615 & 2357 & 1284 \\ 
  DE & 1.00 & 1.01 & 8444 & 8762 & 6229 & 4889 & 1959 \\ 
  Div II & 1.00 & 1.01 & 6235 & 5983 & 5122 & 3523 & 2497 \\ 
  Div III & 1.00 & 1.00 & 7153 & 6796 & 5125 & 5269 & 4765 \\ 
  FBS & 1.00 & 1.00 & 5887 & 7956 & 8008 & 4447 & 4911 \\ 
  FCS & 1.00 & 1.00 & 5039 & 6780 & 5138 & 2211 & 3384 \\ 
  FL & 1.00 & 1.01 & 10521 & 9432 & 4106 & 2971 & 4068 \\ 
  GA & 1.00 & 1.01 & 10256 & 7518 & 4586 & 4261 & 4521 \\ 
  HI & 1.00 & 1.01 & 6318 & 6500 & 3740 & 1726 & 1453 \\ 
  IA & 1.00 & 1.00 & 6783 & 5936 & 4195 & 6483 & 6422 \\ 
  ID & 1.00 & 1.01 & 9125 & 7599 & 4284 & 2923 & 2007 \\ 
  IL & 1.00 & 1.01 & 10734 & 7242 & 4384 & 2663 & 3491 \\ 
  IN & 1.00 & 1.00 & 6848 & 6028 & 4481 & 6731 & 6370 \\ 
  KS & 1.00 & 1.00 & 12518 & 8315 & 5064 & 4064 & 3711 \\ 
  KY & 1.00 & 1.01 & 7930 & 5791 & 4690 & 5257 & 2594 \\ 
  LA & 1.00 & 1.01 & 10124 & 6418 & 5027 & 4489 & 4163 \\ 
  MA & 1.00 & 1.00 & 11382 & 7744 & 3594 & 3393 & 5556 \\ 
  MD & 1.00 & 1.01 & 5596 & 5298 & 4573 & 2887 & 1612 \\ 
  ME & 1.00 & 1.02 & 4482 & 4546 & 4293 & 2581 & 872 \\ 
  MI & 1.00 & 1.00 & 7510 & 6551 & 3462 & 4747 & 4958 \\ 
  MN & 1.00 & 1.00 & 6395 & 5716 & 4130 & 6042 & 6621 \\ 
  MO & 1.00 & 1.00 & 10748 & 7811 & 4349 & 3718 & 4125 \\ 
  MS & 1.00 & 1.00 & 8342 & 5954 & 4480 & 5211 & 5344 \\ 
  MT & 1.00 & 1.01 & 6285 & 5458 & 5068 & 4137 & 929 \\ 
  NC & 1.00 & 1.00 & 12694 & 8651 & 4418 & 4989 & 4031 \\ 
  ND & 1.00 & 1.01 & 8022 & 6847 & 6650 & 3898 & 1562 \\ 
  NE & 1.00 & 1.01 & 6339 & 5449 & 4962 & 4102 & 1477 \\ 
  NFL & 1.00 & 1.00 & 16221 & 8515 & 9909 & 4960 & 10489 \\ 
  NH & 1.00 & 1.01 & 7250 & 6835 & 5240 & 3398 & 1315 \\ 
  NJ & 1.00 & 1.00 & 8816 & 6984 & 4633 & 6478 & 6141 \\ 
  NM & 1.00 & 1.01 & 6466 & 6191 & 2388 & 1360 & 810 \\ 
  NV & 1.00 & 1.01 & 4676 & 4969 & 5626 & 3256 & 920 \\ 
  NY & 1.00 & 1.01 & 9052 & 7607 & 5950 & 4414 & 2310 \\ 
  OH & 1.00 & 1.01 & 11363 & 8122 & 3627 & 2325 & 4370 \\ 
  OK & 1.00 & 1.00 & 9426 & 7070 & 4776 & 5564 & 5807 \\ 
  OR & 1.00 & 1.01 & 6590 & 5797 & 4555 & 3176 & 1669 \\ 
  PA & 1.00 & 1.01 & 10046 & 7744 & 4642 & 4042 & 3402 \\ 
  RI & 1.00 & 1.00 & 10951 & 7570 & 6417 & 3664 & 2457 \\ 
  SC & 1.00 & 1.01 & 5129 & 5442 & 4138 & 1603 & 1969 \\ 
  SD & 1.00 & 1.02 & 5751 & 5906 & 3916 & 2161 & 915 \\ 
  TN & 1.00 & 1.00 & 8425 & 6462 & 4822 & 5623 & 5185 \\ 
  TX & 1.00 & 1.00 & 9465 & 7391 & 3397 & 1665 & 3831 \\ 
  UT & 1.00 & 1.01 & 6343 & 6338 & 4588 & 1425 & 1452 \\ 
  VA & 1.00 & 1.01 & 6359 & 5159 & 4168 & 3229 & 3175 \\ 
  VT & 1.00 & 1.01 & 10132 & 6763 & 5398 & 4005 & 2249 \\ 
  WA & 1.00 & 1.01 & 6106 & 6287 & 3729 & 4285 & 3832 \\ 
  WI & 1.00 & 1.00 & 14213 & 8640 & 5481 & 4388 & 4304 \\ 
  WV & 1.00 & 1.01 & 6372 & 6294 & 4133 & 2456 & 2043 \\ 
  WY & 1.00 & 1.01 & 6036 & 4595 & 4902 & 3311 & 990 \\ 
   \hline
\end{tabular}
\caption{Model diagnostics for Model 1. Minimum and maximum $\hat R$ and effective sample size (ESS) for parameters of interest are shown. For parameters which are vectors in each league (such as $\theta_k$), mean ESS are shown. }\label{tab:s1}
\end{table}

\begin{table}[ht]
\centering
\begin{tabular}{cccccccccc}
  \hline
 League & $\min(\hat R)$ & $\max(\hat R)$ & $\beta_{0k}$ & $\beta_{1k}$ & $\lambda_{0k}$ & $\lambda_{1k}$ & $\sigma_k$ & $\zeta_k$ & $\theta_k$ \\
  \hline
  
AK & 1.00 & 1.01 & 1859 & 3760 & 2421 & 6318 & 3776 & 2488 & 1575 \\ 
  AL & 1.00 & 1.01 & 6793 & 6474 & 9089 & 6848 & 3469 & 2380 & 4024 \\ 
  AR & 1.00 & 1.01 & 3050 & 2875 & 5124 & 5686 & 3800 & 1904 & 2336 \\ 
  AZ & 1.00 & 1.01 & 3190 & 3164 & 6980 & 6855 & 4548 & 1366 & 1248 \\ 
  CA & 1.00 & 1.01 & 5130 & 4950 & 7262 & 6715 & 3246 & 2811 & 3063 \\ 
  CO & 1.00 & 1.00 & 6023 & 5883 & 7870 & 6332 & 4108 & 3304 & 3176 \\ 
  CT & 1.00 & 1.01 & 2718 & 2719 & 4697 & 4713 & 4756 & 2408 & 1378 \\ 
  DE & 1.00 & 1.02 & 2516 & 2412 & 3842 & 5400 & 5074 & 2149 & 1020 \\ 
  Div II & 1.00 & 1.01 & 2890 & 3025 & 6197 & 5468 & 6058 & 3291 & 2324 \\ 
  Div III & 1.00 & 1.00 & 6420 & 6378 & 8163 & 7450 & 5219 & 4352 & 4114 \\ 
  FBS & 1.00 & 1.00 & 2969 & 3070 & 7580 & 6458 & 6580 & 3250 & 4616 \\ 
  FCS & 1.00 & 1.00 & 3112 & 2965 & 6819 & 6302 & 5981 & 2961 & 4157 \\ 
  FL & 1.00 & 1.00 & 5459 & 4900 & 7970 & 6422 & 4155 & 2695 & 4111 \\ 
  GA & 1.00 & 1.01 & 5706 & 5278 & 8784 & 6661 & 4472 & 5396 & 4691 \\ 
  HI & 1.00 & 1.01 & 2856 & 2807 & 5844 & 5846 & 3085 & 1343 & 1257 \\ 
  IA & 1.00 & 1.01 & 5336 & 5502 & 6340 & 5560 & 5077 & 4256 & 3574 \\ 
  ID & 1.00 & 1.01 & 3385 & 3147 & 5978 & 6498 & 4213 & 2522 & 1697 \\ 
  IL & 1.00 & 1.01 & 4791 & 4717 & 7283 & 6040 & 4500 & 3439 & 3628 \\ 
  IN & 1.00 & 1.00 & 7786 & 7500 & 8561 & 6417 & 4935 & 3218 & 3489 \\ 
  KS & 1.00 & 1.01 & 2696 & 2670 & 4952 & 5204 & 4252 & 1921 & 1672 \\ 
  KY & 1.00 & 1.01 & 3815 & 3758 & 7112 & 6170 & 5590 & 4494 & 2340 \\ 
  LA & 1.00 & 1.00 & 6626 & 6121 & 7542 & 6018 & 4835 & 4856 & 4356 \\ 
  MA & 1.00 & 1.00 & 6653 & 6898 & 8034 & 5992 & 4372 & 3251 & 5340 \\ 
  MD & 1.00 & 1.01 & 3245 & 3086 & 5969 & 5245 & 3866 & 4328 & 1995 \\ 
  ME & 1.00 & 1.01 & 2165 & 2215 & 5185 & 6418 & 4575 & 2928 & 1050 \\ 
  MI & 1.00 & 1.01 & 4975 & 4653 & 7218 & 6650 & 3892 & 3262 & 3947 \\ 
  MN & 1.00 & 1.01 & 3827 & 3686 & 6179 & 5614 & 4799 & 3856 & 3263 \\ 
  MO & 1.00 & 1.00 & 6115 & 5911 & 6777 & 6264 & 4676 & 4311 & 5102 \\ 
  MS & 1.00 & 1.01 & 4849 & 5135 & 6621 & 5852 & 4310 & 3413 & 3629 \\ 
  MT & 1.00 & 1.01 & 2575 & 2561 & 6124 & 6466 & 5203 & 3827 & 938 \\ 
  NC & 1.00 & 1.00 & 8058 & 7695 & 8990 & 6570 & 4753 & 5931 & 3974 \\ 
  ND & 1.00 & 1.01 & 3009 & 2996 & 6695 & 5887 & 5062 & 3523 & 1387 \\ 
  NE & 1.00 & 1.01 & 3073 & 2873 & 5362 & 6333 & 4798 & 3344 & 1198 \\ 
  NFL & 1.00 & 1.00 & 6639 & 6562 & 8633 & 7641 & 10177 & 4286 & 9451 \\ 
  NH & 1.00 & 1.01 & 2505 & 2531 & 5334 & 5358 & 4073 & 2356 & 1173 \\ 
  NJ & 1.00 & 1.00 & 7735 & 7579 & 7780 & 6014 & 5001 & 4996 & 4938 \\ 
  NM & 1.00 & 1.02 & 2798 & 2792 & 5791 & 5899 & 2808 & 1174 & 796 \\ 
  NV & 1.00 & 1.01 & 3152 & 3110 & 5007 & 5879 & 4552 & 2562 & 1152 \\ 
  NY & 1.00 & 1.01 & 4162 & 4503 & 6785 & 6817 & 6333 & 4382 & 1985 \\ 
  OH & 1.00 & 1.01 & 4926 & 4807 & 8341 & 6298 & 4591 & 2424 & 3675 \\ 
  OK & 1.00 & 1.01 & 4681 & 4722 & 5916 & 5534 & 4659 & 4624 & 3951 \\ 
  OR & 1.00 & 1.01 & 2664 & 2637 & 4974 & 5275 & 4305 & 2800 & 1650 \\ 
  PA & 1.00 & 1.01 & 3649 & 3522 & 7878 & 6796 & 4651 & 4098 & 2848 \\ 
  RI & 1.00 & 1.01 & 2825 & 2823 & 5540 & 5668 & 3898 & 1976 & 1522 \\ 
  SC & 1.00 & 1.01 & 3220 & 3249 & 5432 & 5061 & 4303 & 1428 & 1757 \\ 
  SD & 1.00 & 1.02 & 2830 & 2848 & 4943 & 5897 & 4070 & 2333 & 891 \\ 
  TN & 1.00 & 1.00 & 7370 & 6841 & 7166 & 5649 & 4925 & 5835 & 5177 \\ 
  TX & 1.00 & 1.01 & 3843 & 3833 & 5784 & 6639 & 3197 & 1613 & 3308 \\ 
  UT & 1.00 & 1.02 & 1703 & 2033 & 5577 & 4966 & 2646 & 1217 & 1134 \\ 
  VA & 1.00 & 1.00 & 6745 & 6875 & 6600 & 5949 & 5259 & 5605 & 4719 \\ 
  VT & 1.00 & 1.00 & 3545 & 3781 & 6298 & 7149 & 5046 & 3605 & 2115 \\ 
  WA & 1.00 & 1.00 & 5357 & 5680 & 7110 & 5814 & 5069 & 5379 & 5124 \\ 
  WI & 1.00 & 1.01 & 4728 & 4574 & 6616 & 6620 & 5390 & 3110 & 3101 \\ 
  WV & 1.00 & 1.01 & 2437 & 2454 & 5592 & 5486 & 3819 & 2186 & 1682 \\ 
  WY & 1.00 & 1.01 & 4131 & 4057 & 4799 & 5630 & 5474 & 4813 & 1416 \\ 
   \hline
\end{tabular}
\caption{Model diagnostics for Model 2. Minimum and maximum $\hat R$ and effective sample size (ESS) for parameters of interest are shown. For parameters which are vectors in each league (such as $\theta_k$), mean ESS are shown. }\label{tab:s2}
\end{table}

\begin{table}[ht]
\centering
\begin{tabular}{cccccccc}
  \hline

League & $\min(\hat R)$ & $\max(\hat R)$ & $\gamma_{kt}$ & $\tau_k$ & $\sigma_k$ & $\zeta_k$ & $\theta_k$  \\
  \hline
AK & 1.00 & 1.00 & 6435 & 1108 & 4827 & 4470 & 3825 \\ 
  AL & 1.00 & 1.01 & 9208 & 5819 & 3778 & 2254 & 3256 \\ 
  AR & 1.00 & 1.01 & 5885 & 4100 & 4180 & 1484 & 2195 \\ 
  AZ & 1.00 & 1.01 & 5865 & 4550 & 3585 & 2607 & 3052 \\ 
  CA & 1.00 & 1.01 & 8700 & 6981 & 3411 & 2592 & 3093 \\ 
  CO & 1.00 & 1.01 & 5935 & 4133 & 3316 & 2242 & 1924 \\ 
  CT & 1.00 & 1.02 & 4942 & 3758 & 4935 & 1907 & 1259 \\ 
  DE & 1.00 & 1.01 & 6515 & 1389 & 5634 & 4171 & 1632 \\ 
  Div II & 1.00 & 1.01 & 6412 & 4486 & 5201 & 4194 & 2968 \\ 
  Div III & 1.00 & 1.01 & 6098 & 4990 & 4860 & 3126 & 2426 \\ 
  FBS & 1.00 & 1.00 & 7382 & 5785 & 5710 & 3688 & 4706 \\ 
  FCS & 1.00 & 1.00 & 6703 & 4875 & 4554 & 2639 & 3615 \\ 
  FL & 1.00 & 1.00 & 11243 & 6704 & 4500 & 3675 & 4140 \\ 
  GA & 1.00 & 1.00 & 12237 & 8003 & 5215 & 5162 & 4660 \\ 
  HI & 1.00 & 1.01 & 5561 & 2282 & 4050 & 1651 & 1413 \\ 
  IA & 1.00 & 1.00 & 8621 & 5487 & 4735 & 6142 & 4927 \\ 
  ID & 1.00 & 1.01 & 7361 & 2570 & 3839 & 2825 & 1936 \\ 
  IL & 1.00 & 1.00 & 7967 & 5767 & 4284 & 3827 & 4759 \\ 
  IN & 1.00 & 1.00 & 12961 & 6592 & 4090 & 3260 & 3695 \\ 
  KS & 1.00 & 1.01 & 6632 & 3945 & 4174 & 2767 & 2275 \\ 
  KY & 1.00 & 1.01 & 6396 & 5015 & 5625 & 3293 & 1806 \\ 
  LA & 1.00 & 1.00 & 10279 & 6107 & 5381 & 5392 & 5127 \\ 
  MA & 1.00 & 1.01 & 6802 & 4987 & 3786 & 2800 & 3655 \\ 
  MD & 1.00 & 1.01 & 4893 & 3521 & 4683 & 3089 & 1567 \\ 
  ME & 1.00 & 1.01 & 4469 & 2456 & 3911 & 3142 & 986 \\ 
  MI & 1.00 & 1.01 & 10131 & 6664 & 4652 & 3184 & 4300 \\ 
  MN & 1.00 & 1.00 & 6935 & 5007 & 4723 & 5470 & 5545 \\ 
  MO & 1.00 & 1.00 & 11317 & 7161 & 4173 & 3429 & 4952 \\ 
  MS & 1.00 & 1.00 & 7986 & 5546 & 4608 & 4682 & 5552 \\ 
  MT & 1.00 & 1.01 & 6331 & 3889 & 5463 & 4136 & 1003 \\ 
  NC & 1.00 & 1.00 & 12466 & 7904 & 4876 & 5700 & 4334 \\ 
  ND & 1.00 & 1.01 & 6832 & 3643 & 5662 & 3968 & 1605 \\ 
  NE & 1.00 & 1.02 & 5004 & 2460 & 5278 & 3567 & 1170 \\ 
  NFL & 1.00 & 1.00 & 13948 & 9185 & 11446 & 5404 & 10783 \\ 
  NH & 1.00 & 1.01 & 6573 & 2556 & 4850 & 3253 & 1338 \\ 
  NJ & 1.00 & 1.00 & 8924 & 7308 & 4446 & 6428 & 5786 \\ 
  NM & 1.00 & 1.02 & 5047 & 2971 & 3359 & 1235 & 713 \\ 
  NV & 1.00 & 1.01 & 6356 & 3330 & 5298 & 4098 & 1126 \\ 
  NY & 1.00 & 1.01 & 8957 & 6528 & 4834 & 4649 & 2939 \\ 
  OH & 1.00 & 1.00 & 11147 & 7157 & 4266 & 3740 & 4444 \\ 
  OK & 1.00 & 1.00 & 8430 & 6473 & 4499 & 6056 & 5391 \\ 
  OR & 1.00 & 1.01 & 5077 & 3706 & 4201 & 2586 & 1444 \\ 
  PA & 1.00 & 1.01 & 9493 & 6421 & 4766 & 4010 & 3340 \\ 
  RI & 1.00 & 1.01 & 6778 & 1665 & 6143 & 3644 & 2331 \\ 
  SC & 1.00 & 1.01 & 5689 & 4400 & 4729 & 2947 & 2520 \\ 
  SD & 1.00 & 1.01 & 6989 & 2929 & 4426 & 2821 & 1106 \\ 
  TN & 1.00 & 1.00 & 13272 & 9891 & 5095 & 5730 & 4583 \\ 
  TX & 1.00 & 1.01 & 8695 & 5901 & 3388 & 1747 & 3723 \\ 
  UT & 1.00 & 1.01 & 6149 & 4077 & 3470 & 1199 & 1426 \\ 
  VA & 1.00 & 1.00 & 9841 & 5726 & 4095 & 5583 & 4084 \\ 
  VT & 1.00 & 1.00 & 7691 & 4758 & 4837 & 4315 & 2927 \\ 
  WA & 1.00 & 1.01 & 6262 & 4508 & 4518 & 4802 & 4143 \\ 
  WI & 1.00 & 1.00 & 12594 & 7848 & 4706 & 4080 & 4132 \\ 
  WV & 1.00 & 1.01 & 6015 & 2630 & 4373 & 2711 & 1937 \\ 
  WY & 1.00 & 1.01 & 6813 & 3389 & 4961 & 3772 & 1179 \\ 
   \hline
\end{tabular}
\caption{Model diagnostics for Model 3. Minimum and maximum $\hat R$ and effective sample size (ESS) for parameters of interest are shown. For parameters which are vectors in each league (such as $\theta_k$ and  $\gamma_{kt}$), mean ESS are shown. }\label{tab:s3}
\end{table}

\begin{table}[ht]
\centering
\begin{tabular}{ccccccccccc}
  \hline
Model & $\min(\hat R)$ & $\max(\hat R)$ & $\beta_{0k}$ & $\beta_{1k}$ & $\beta_1^*$ &  $\lambda_{0k}$& $\lambda_1$ & $\sigma_k$ & $\zeta_k$ & $\theta_k$\\ 
  \hline
  States $<$ 10,000 Games & 1.00 & 1.02 & 1987 & 1384 & 1050 & 4564 & 313 & 3069 & 2005 & 1079 \\ 
All 50 States & 1.00 & 1.02 & 4240 & 2319 & 1332 & 7134 & 115 & 5074 & 3257 & 2847 \\ 
   \hline
\end{tabular}
\caption{Model diagnostics for hierarchical models. Minimum and maximum $\hat R$ and effective sample size (ESS) for parameters of interest are shown. For parameters which are vectors in each model, mean ESS are shown.}
\label{tab:s4}
\end{table}